
\def\CC{{\mathchoice
{\rm C\mkern-8mu\vrule height1.45ex depth-.05ex 
width.05em\mkern9mu\kern-.05em}
{\rm C\mkern-8mu\vrule height1.45ex depth-.05ex 
width.05em\mkern9mu\kern-.05em}
{\rm C\mkern-8mu\vrule height1ex depth-.07ex 
width.035em\mkern9mu\kern-.035em}
{\rm C\mkern-8mu\vrule height.65ex depth-.1ex 
width.025em\mkern8mu\kern-.025em}}}

\def\RR{{\rm I\kern-1.6pt {\rm R}}}

\def\ZZ{{\rm Z}\kern-3.8pt {\rm Z} \kern2pt}

\input phyzzx.tex

\def\np{Nucl. Phys.}
\def\pl{Phys. Lett.}

\def\cmp{Comm. Math. Phys.}
\def\jmp{J. Math. Phys.}
\def\ijmp{Int. J. Mod. Phys.}
\def\mpl{Mod. Phys. Lett.}
\def\lmp{Lett. Math. Phys.}

\def\adm{Adv. in Math.}

\def\fap{Funkt. Anal. Prilozhen}
\def\pnas{Proc. Natl. Acad. Sci. USA}
\def\sjnp{Sov. J. Nucl. Phys.}

\def\qjm{Quart. J. Math. Oxford Ser. (2)}

\tolerance=500000
\overfullrule=0pt
\pubnum={US-FT-12/97\cr hep-th/9704065}
\date={April, 1997}
\pubtype={}
\titlepage

\title{Fusion rules and singular vectors of the 
${\rm osp}(1\vert 2)$ 
current algebra} 
\author{I.P. Ennes\foot{E-mail: ENNES@GAES.USC.ES} and 
A.V. Ramallo\foot{E-mail: ALFONSO@GAES.USC.ES} }
\address{Departamento de F\'\i sica de
Part\'\i culas, \break Universidad de Santiago, \break
E-15706 Santiago de Compostela, Spain. \break}

\abstract{The fusion of Verma modules of the 
${\rm osp}(1\vert 2)$ current algebra is studied. 
In the framework of an isotopic formalism, the singular
vector decoupling conditions are analyzed. The fusion
rules corresponding to the admissible representations of
the ${\rm osp}(1\vert 2)$ algebra are determined. A
relation between the characters of these last
representations and those corresponding to the minimal
superconformal models is found. A series of equations
that relate the descendants of the highest weight vectors
resulting from a  fusion of Verma modules are obtained.
Solving these equations the singular vectors of the
theory can be determined.}

\endpage
\pagenumber=1

\chapter{Introduction}

The models endowed with a current algebra symmetry based
on an affine Lie algebra form a distinguished class in 
Conformal Field Theory (CFT). Indeed, in recent years a
rich variety of results has been obtained for this class
of models 
\REF\Review{For a review see J. Fuchs, ``Affine Lie algebras
and quantum groups", Cambridge University Press, 1992 and 
S. Ketov, ``Conformal Field Theory", World Scientific,
Singapore(1995).}
[\Review]. Most of these results correspond to
the so-called integrable representations, which appear when
the level $k$ of the affine symmetry is a positive 
integer. It turns out, however, that a larger class of
representations and levels is needed in order to
relate the model based on the current algebra to other
CFT's that do not enjoy this symmetry. Of particular
relevance are the admissible representations 
\REF\KW{V. Kac and M. Wakimoto\journal\pnas&85(88)4956.}
[\KW] which,
for example, in the case of the $sl(2)$ algebra, are those
needed to construct the minimal Virasoro models by means
of the hamiltonian reduction procedure 
\REF\BO{M. Bershadsky and H. Ooguri
\journal\cmp&126(89)49.}
[\BO]. These
admissible representations, which in the $sl(2)$ case
appear when the level and the isospin are rational
numbers, are also necessary in the light-cone analysis of
the two dimensional quantum gravity coupled to minimal
matter 
\REF\KPZ{A. M. Polyakov\journal\mpl&A2(87)893; V. G.
Knizhnik, A. M. Polyakov and A. B.
Zamolodchikov\journal\mpl&A3(88)819.}
[\KPZ] and in the study of the non-critical string
theories by means of the topological coset models 
\REF\hu{H.L. Hu and M. Yu \journal\pl&B289(92)302
\journal\np&B391(93)389.}
\REF\yank{M. Spiegelglas and S. Yankielowicz
\journal\np&393(93)301;  O. Aharony et al.\journal
\np&B399(93)527\journal\pl&B289(92)309
\journal\pl&B305(93)35.}
[\hu,\yank].

It is therefore interesting to extend the concepts and
techniques of CFT to include  general types of
levels and representations. In the $sl(2)$ case, this
problem has been addressed following different
approaches 
\REF\MFF{F. G. Malikov, B. L. Feigin and D. B.
Fuks\journal\fap&20,No. 2(86)25.}
\REF\BF{D. Bernard and G. Felder\journal\cmp&127(90)145.}
\REF\Feigin{B. L. Feigin and E. V. Frenkel in ``Physics and
Mathematics of Strings", edited by L. Brink et al., World
Scientific, 1990 \journal\cmp&128(90)161 
\journal\lmp&19(90)307.} 
\REF\Furlan{P. Furlan, A. Ch. Ganchev, R. Paunov and V.
B. Petkova\journal\pl&B267(91)63\journal\np&B394(93)665; 
A. Ch. Ganchev and V.B. Petkova\journal\pl&B293(92)56.}
\REF\AY{H. Awata and Y. Yamada\journal\mpl&A7(92)1185.}
\REF\Bauer{N. Bauer and N. Sochen\journal\cmp&152(93)127.}
\REF\peter{J. L. Petersen, J. Rasmussen and M. Yu
\journal\np&B457(95)309\journal\np&B457(95)343.}
\REF\Andreev{O. Andreev\journal\pl&B363(95)166.}
[\MFF -\Andreev]. The key point in most of these analysis
is the introduction of an isotopic variable
$x$ to represent the $sl(2)$ symmetry 
\REF\FZ{A. B. Zamolodchikov and V. A.
Fateev\journal\sjnp&43(86)657.}
[\FZ]. The primary
fields of the theory (and thus its conformal blocks)
depend both on the spacetime coordinate $z$ and on the
``internal" coordinate
$x$. Within this isotopic formalism it has been possible
[\Bauer ] to give a precise definition of the fusion of
primary fields. This  has allowed to develop an
efficient algorithm to compute the singular vectors of the
$sl(2)$ affine algebra. Moreover, in refs. 
\REF\FM{B. Feigin and F. Malikov\journal\lmp&31(94)315;
``Modular functor and representation theory at  
$\hat sl_2$ at a rational level", q-alg/9511011. }
[\AY, \FM] it
has been shown that the primary fields corresponding to the
admissible representations close a well-defined fusion
algebra. By using a free field realization of the current
algebra (supplemented with the fractional calculus
technique 
\REF\Fractional{A. C. Mc. Bride and G. F. Roach(eds),
``Fractional Calculus"  (Pitman Advanced Publishing
Program, Boston, 1985); S. G. Samko, A. A. Kilbas and O.
L. Marichec, ``Fractional integrals and derivatives"
(Gordon and Breach Science Publishers, 1993).}
[\Fractional]), one can represent [\peter] the
conformal blocks for the correlators of primary fields
which carry quantum numbers of $sl(2)$ admissible
representations (\ie\ with fractional isospins). When the
isotopic and space-time coordinates are identified, the
quantum hamiltonian reduction is implemented and one
passes from the blocks of the $sl(2)$ theory to those of
the minimal Virasoro models [\Furlan].

In this paper we shall study the ${\rm osp}(1\vert 2)$
affine Lie superalgebra 
\REF\Pais{A. Pais and V. Rittenberg\journal\jmp&16(75)2063; 
M. Scheunert, W. Nahn and  V. Rittenberg\journal\jmp&18(77)155.}
\REF\Kac{V. G.
Kac\journal\adm&26(77)8\journal\adm&30(79)85.}
[\Pais, \Kac]. The r\^ole of this
superalgebra, which is a graded version of $sl(2)$ 
\REF\scheunert{For a review on general aspects of Lie
superalgebras see 
M. Scheunert,  ``The Theory of Lie
Superalgebras",  {\sl Lect. Notes in Math.} 716,
Springer-Verlag, Berlin (1979) and  
L. Frappat, P. Sorba and A. Sciarrino, ``Dictionary on
Lie Superalgebras", hep-th/9607161.}
[\scheunert], in connection with the N=1 superconformal
symmetry is well-known. In fact, by means of the
hamiltonian reduction method, the ${\rm osp}(1\vert 2)$
CFT can be related to the minimal N=1 superconformal
models 
\REF\bershadsky{M. Bershadsky and H.
Ooguri\journal\pl&B229(89)374.}
[\bershadsky]. The 
${\rm osp}(1\vert 2)$ algebra also appears in the
light-cone quantization of two-dimensional supergravity
\REF\poly{A. M. Polyakov and A. B.
Zamolodchikov\journal\mpl&A3(88)1213.}
[\poly]  and  the corresponding topological coset, \ie\
the 
${\rm osp}(1\vert 2)/{\rm osp}(1\vert 2)$ model, can be
used to describe the non-critical Ramond-Neveu-Schwarz
superstrings
\REF\yu{J. B. Fan and M. Yu, ``G/G Gauged
Supergroup Valued WZNW Field Theory", Academia Sinica preprint
AS-ITP-93-22, hep-th/9304123.}
\REF\Ennes{I. P. Ennes, J. M. Isidro and A. V.
Ramallo\journal\ijmp&A11(96)2379.}
[\yu, \Ennes].

For integer level and integer or half-integer isospins,
the ${\rm osp}(1\vert 2)$ conformal blocks have been
studied in ref. 
\REF\osp{I. P. Ennes, A. V. Ramallo and J. M. Sanchez de
Santos\journal\pl&B389(96)485; 
 ``Structure constants for the
${\rm osp}(1\vert 2)$ current algebra", Santiago preprint
US-FT-40/96, hep-th/9610224, to appear in Nucl. Phys. B.}
[\osp]. For more general representations
and levels one has to work with an isotopic representation
which, in addition to a bosonic internal coordinate that
also appears in the $sl(2)$ case, requires the
introduction of an internal Grassmann odd coordinate
$\theta$. Following the approach of ref. [\Bauer], we shall
be able to characterize, in this isotopic formalism, the
fusion of ${\rm osp}(1\vert 2)$ primary fields. This
fusion is greatly constrained when there are singular
vectors in any of the Verma modules associated to the
primary fields participating in the fusion. Actually, we
shall find a set of polynomial relations for the isospins
of these primary fields that encode the decoupling of the
singular vectors in the highest weight module in which
they are originated. These relations are enough to
determine the fusion rules satisfied by the admissible
representations. In analogy with what happens with the
$sl(2)$ algebra [\AY, \FM], we shall find two types of
fusion rules which cannot be simultaneously satisfied. In
these fusion rules the operator algebra closes when the
primary fields are restricted to belong to a conformal
grid, which constitutes a new similarity between this kind
of representations and those corresponding to the minimal
N=1 superconformal models.

The descendants of the highest weight vectors resulting
from a fusion of primary fields satisfy the so-called
descent equations [\Bauer]. These equations can be
reformulated, by means of the Sugawara construction of the
energy-momentum tensor, in such a way that they can be
put in a triangular form, which provides a method to
compute the descendant vectors. As was the case for the
$sl(2)$ algebra [\Bauer], the solution of the descent
equations will allow us to develop an algorithm for the
computation of the singular vectors of the algebra.

The organization of this paper is the following. In
section 2 we give the basic definitions and set up the
formalism for the fusion of ${\rm osp}(1\vert 2)$ Verma
modules. The decoupling conditions induced in the fusion
by the presence of a singular vector are worked out in
section 3. In section 4 we restrict ourselves to the case
of admissible representations. The fusion rules
corresponding to these representations are found as a
consequence of the decoupling conditions obtained in
section 3. The descent equations are derived in section
5. The solution of these equations and their truncation
are also analyzed in this section. The Sugawara recursion
relations are obtained in section 6, where it is also
shown how to use them  to calculate
singular vectors of the algebra. In section 7 we rederive
the recursion relations of section 6 from the
Knizhnik-Zamolodchikov equation. The results obtained in
the paper are summarized in section 8, where some
possible lines of future work are mentioned.

The paper ends with three appendices. In appendix A, the
calculation of the characters of the ${\rm osp}(1\vert 2)$
affine algebra is reviewed and a relation between the
characters of the ${\rm osp}(1\vert 2)$ admissible
representations and those of the minimal superconformal
models is obtained. In appendix B, the two simplest
singular vectors of the algebra are computed by using the
fusion formalism developed in the main text. Finally, in
appendix C, the equivalence of two vectors, needed in the
derivation of section 7, is proved.

\chapter{Fusion of Verma modules}

The ${\rm osp}(1\vert 2)$ current algebra (which we shall 
denote simply by ${\cal A}$) is generated by the currents 
$J_n^{a}$ ($a=0,\pm$, $n\in\ZZ$) and 
$j_n^{\alpha}$ ($\alpha=\pm$, $n\in\ZZ$), together with a
central element $k$. The $J_n^{a}$( $j_n^{\alpha}$)
currents are bosonic (fermionic), \ie\ Grassmann even
(odd). The generator $k$ commutes with all the other
elements of ${\cal A}$ and therefore we shall regard it as
a c-number (the level of ${\cal A}$). Notice that 
the modes $n$ of 
the fermionic currents $j_n^{\alpha}$ run over the 
integers, which,
properly speaking, means that we are considering the
Ramond sector of the ${\rm osp}(1\vert 2)$ affine
superalgebra. The non-vanishing (anti)commutators of 
${\cal A}$ are:
$$
\eqalign{
&[\,J_n^0\,,\,J_m^{\pm}\,]\,=\,\pm J_{n+m}^{\pm}
\,\,\,\,\,\,\,\,\,\,\,\,\,\,\,\,\,\,\,\,\,\,\,\,\,
[\,J_n^0\,,\,J_m^{0}\,]\,=\,{k\over 2}\,n\,\delta_{n+m}\cr
&[\,J_n^+\,,\,J_m^{-}\,]\,=\,kn\delta_{n+m}\,+\,
2J^0_{n+m}\cr
&[\,J_n^0\,,\,j_m^{\pm}\,]\,=\,\pm\,{1\over 2}\,
j_{m+n}^{\pm}
\,\,\,\,\,\,\,\,\,\,\,\,\,\,\,\,\,\,\,\,\,\,\,\,\,
[\,J_n^{\pm}\,,\,j_m^{\pm}\,]\,=\,0\cr
&[\,J_n^{\pm}\,,\,j_m^{\mp}\,]\,=\,-j_{n+m}^{\pm}
\,\,\,\,\,\,\,\,\,\,\,\,\,\,\,\,\,\,\,\,\,\,\,\,\,
\{\,j_n^{\pm}\,,\,j_m^{\pm}\,\}\,=\,\pm 2 J_{n+m}^{\pm}\cr
&\{\,j_n^+\,,\,j_m^{-}\,\}\,=\,2kn\delta_{n+m}\,+\,
2J^0_{n+m}\,\,.\cr}
\eqn\uno
$$
It is interesting to point out that this algebra is
doubly graded. Let us denote by $\overline{d}$ and 
$\underline{d}$ the corresponding gradations. They are
defined by:
$$
\eqalign{
&\overline{d}(\,J_n^a\,)\,=\,a
\,\,\,\,\,\,\,\,\,\,\,\,\,\,\,
\overline{d}(\,j_n^{\alpha}\,)\,=\,{\alpha\over 2}
\,\,\,\,\,\,\,\,\,\,\,\,\,\,\,
\overline{d}(\,k\,)\,=\,0\cr
&\underline{d}(\,J_n^a\,)\,=\,n
\,\,\,\,\,\,\,\,\,\,\,\,\,\,\,
\underline{d}(\,j_n^{\alpha}\,)\,=\,n
\,\,\,\,\,\,\,\,\,\,\,\,\,\,\,
\underline{d}(\,k\,)\,=\,0\,\,.\cr}
\eqn\dos
$$
From the commutation relations \uno\ it follows that 
$\overline{d}$ can be represented as $ad(\,J_0^0\,)$. For
this reason we shall refer to $\overline{d}$ as the 
$J_0^0$-gradation. The so-called principal gradation
\REF\Kacbook{V. G. Kac, ``Infinite dimensional Lie
algebras", Cambridge University Press, 1985.}
[\Kacbook] $d$ is obtained by combining $\overline{d}$ 
and 
$\underline{d}$ in the form:
$$
d\,=\,2\,\underline{d}\,+\,\overline{d}\,\,.
\eqn\tres
$$
Using eqs. \dos\ and \tres\ one can immediately
obtain the $d$-grading of the different generators of 
${\cal A}$:
$$
d(\,J_n^a\,)\,=\,2n\,+\,a
\,\,\,\,\,\,\,\,\,\,\,\,\,\,\,
d(\,j_n^{\alpha}\,)\,=\,2n\,+\,{\alpha\over 2}
\,\,\,\,\,\,\,\,\,\,\,\,\,\,\,
d(\,k\,)\,=\,0\,\,.
\eqn\cuatro
$$
With respect to $d$ the algebra ${\cal A}$ splits as:
$$
{\cal A}\,=\,{\cal A}_-\,\oplus\,{\cal A}_0\,
\oplus\,{\cal A}_+\,\,,
\eqn\cinco
$$
where ${\cal A}_-$,  ${\cal A}_0$ and  ${\cal A}_+$ are
the subspaces of  ${\cal A}$ spanned by the elements that
have, respectively, $d<0$, $d=0$ and $d>0$. These
elements are easy to identify from eq. \cuatro\ and so,
for example,  ${\cal A}_0$ is generated by $J_0^0$ and
$k$, whereas  ${\cal A}_+$ is the subspace spanned by 
$J_n^-\,\,(n\ge 1)$, $J_n^0\,\,(n\ge 1)$, 
$J_n^+\,\,(n\ge 0)$, $j_n^-\,\,(n\ge 1)$ and 
$j_n^+\,\,(n\ge 0)$. 

From the mode operators $J_n^a$ and $j_n^{\alpha}$ one
can define the coordinate-dependent fields $J^{a}(z)$ and
$j^{\alpha}(z)$ as:
$$
J^{a}(z)\,=\,\sum_{n=-\infty}^{+\infty}\,\,
J^{a}_n\,z^{-n-1}
\,\,\,\,\,\,\,\,\,\,\,\,\,\,\,\,\,\,\,\,
j^{\alpha}(z)\,=\,\sum_{n=-\infty}^{+\infty}\,\,
j^{\alpha}_n\,z^{-n-1}\,\,.
\eqn\seis
$$
As it is well-known, one can construct a Conformal Field
Theory associated to the currents $J^{a}(z)$ and
$j^{\alpha}(z)$. The key ingredient in this construction
is the Sugawara prescription for the
energy-momentum tensor $T(z)$ of the theory as a
quadratic expression in the currents. In our case, taking
into account the form of the quadratic Casimir invariant
of ${\rm osp}(1\vert 2)$, one has:
$$
\eqalign{
T(z)\,=\,{1\over 2k+3}\,:\,[\,&2\,(J^0(z))^2\,+\,
J^+(z)\,J^-(z)\,+\,
J^-(z)\,J^+(z)\,-\cr
&-\,{1\over 2}\,j^+(z)\,j^-(z)\,+\,
{1\over 2}\,j^-(z)\,j^+(z)\,]:\,\,.\cr}
\eqn\siete
$$
In eq. \siete\ the double dot $:$ denotes
normal-ordering. The energy-momentum tensor defined in
eq. \siete\ is such that the currents are primary
dimension-one operators with respect to $T(z)$. Its mode
expansion is:
$$
T(z)\,=\,\sum_{n=-\infty}^{+\infty}\,\,
L_n\,z^{-n-2}\,\,,
\eqn\ocho
$$
where the $L_n$'s are the generators of the Virasoro
algebra. The commutators of these operators with the
currents are:
$$
[\,L_n\,,\,J_m^a\,]\,=\,-m\,J_{n+m}^a
\,\,\,\,\,\,\,\,\,\,\,\,\,\,\,\,\,\,\,\,
[\,L_n\,,\,j_m^{\alpha}\,]\,=\,-m\,j_{n+m}^{\alpha}\,\,,
\eqn\nueve
$$
from which it follows that $\underline{d}$ can be
represented as $-ad(\,L_0\,)$, and therefore we shall call 
$\underline{d}$ the $L_0$-gradation. The algebra
satisfied by the  $L_n$'s is the Virasoro algebra, \ie:
$$
[\,L_n,L_m\,]\,=\,(n-m)\,L_{n+m}\,+\,{c\over 12}\,\,
(m^3\,-\,m)\,\delta_{n+m,0}\,\,,
\eqn\diez
$$
where the central charge $c$ is related to the level $k$
by means of the expression:
$$
c\,=\,{2k\over 2k+3}\,\,.
\eqn\once
$$
Instead of the level $k$ we shall frequently use in what
follows the quantity $t$, which in terms of the former is
given by:
$$
t\,=\,2k\,+\,3\,\,.
\eqn\doce
$$

The Verma modules associated to ${\cal A}$ are
constructed by acting with elements of the universal
enveloping algebra of ${\cal A}_-$ (denoted by 
$U(\,{\cal A}_-\,)$) on a highest weight vector $|\,j,t>$.
The latter is annihilated by the elements of  
${\cal A}_+$, \ie:
$$
J_n^a\,|\,j,t>\,=\,j_n^{\alpha}\,|\,j,t>\,=\,0\,\,,
\,\,\,\,\,\,\,\,\,\,\,\,\,\,\,\,\,\,
\forall\,\,(\,J_n^a\,,\,j_n^{\alpha}\,)\,\in\,
{\cal A}_{+}\,\,.
\eqn\trece
$$
On the contrary, $J_0^0$ and $L_0$ act diagonally on 
$|\,j,t>$:
$$
J_0^0\,|\,j,t>\,=\,j\,|\,j,t>
\,\,\,\,\,\,\,\,\,\,\,\,\,\,\,\,\,\,
L_0\,|\,j,t>\,=\,h_j\,|\,j,t>\,\,.
\eqn\catorce
$$
From the Sugawara expression for $L_0$ (see eqs. \siete\
and \ocho), one can easily get the $L_0$ eigenvalue
corresponding to $|\,j,t>$, namely:
$$
h_j\,=\,{j\,(\,2j\,+\,1\,)\over 2k\,+\,3}\,=\,
{\,j(\,2j\,+\,1\,)\over t}\,\,.
\eqn\quince
$$

In order to completely characterize  the highest
weight vector $|\,j,t>$ we must specify its Grassmann
parity, which we shall denote by $p(j)$ : $|\,j,t>$ is
bosonic (fermionic) if $p(j)=0$ ($p(j)=1$). The Verma
module whose highest weight vector is $|\,j,t>$ will be
denoted by $V^{(j,t)}$. Any element in $V^{(j,t)}$ is of
the form 
$u_-|\,j,t>$, where $u_-\in {\cal A}_-$. The gradations 
$\underline{d}$ and $\overline{d}$ of ${\cal A}$ induce a
doubly graded structure in $V^{(j,t)}$. Actually, if we
denote by $n$ and $m$ the eigenvalues of $-\underline{d}$
and $-\overline{d}$ respectively, one can decompose 
$V^{(j,t)}$ as:
$$
V^{(j,t)}\,=\,\bigoplus_{(n,m)}\,V^{(j,t)}_{n,m}\,\,.
\eqn\dseis
$$
Notice that, according to the Poincar\'e-Birkhoff-Witt
theorem, $U(\,{\cal A}_-\,)$ is generated by monomials
and thus we can consider a basis of $V^{(j,t)}$
constituted by vectors of the form:
$$
{|\,\{m_i^a\}\,;j\,>}\,=\,
\prod_{i=0}^{+\infty}\,\Bigl(\,j_{-i}^-\,\Bigr)^{2m_i^-}\,\,
\prod_{i=1}^{+\infty}\,\Bigl(\,J_{-i}^0\,\Bigr)^{m_i^0}\,\,
\prod_{i=1}^{+\infty}\,\Bigl(\,j_{-i}^+\,\Bigr)^{2m_i^+}\,\,
|\,j,t>\,\,.
\eqn\dsiete
$$
In eq. \dsiete, the numbers 
$m_i^{\pm}$ are integers or half-integers whereas the 
$m_i^0$'s are always integers ($m_i^a\ge0$). It is
important to point out that the vectors defined in eq.
\dsiete\ are homogeneous, \ie\ they have  well-defined
$J_0^0$ and $L_0$ eigenvalues. Actually, the $L_0$
eigenvalue  for the vector \dsiete\ is $h_j+n$, where $n$
is given by:
$$
n\,=\,2\,\sum_{i=0}^{+\infty}\,im_i^-\,+\,
\sum_{i=1}^{+\infty}\,im_i^0\,+\,
2\sum_{i=1}^{+\infty}\,im_i^+\,\,,
\eqn\docho
$$
while, if we define $m^{\pm}$ as:
$$
m^+\,=\,\sum_{i=1}^{+\infty}\,m_i^+\,\,,
\,\,\,\,\,\,\,\,\,\,\,\,
m^-\,=\,\sum_{i=0}^{+\infty}\,m_i^-\,\,,
\eqn\dnueve
$$
the value of $m$ for the vector $|\,\{m_i^a\}\,;j\,>$ is
simply:
$$
m\,=\,m^-\,-\,m^+\,\,.
\eqn\veinte
$$
Notice that the vectors \dsiete\ have also a well-defined
Grassmann parity. Actually,  by inspecting eq. \dsiete\
one  easily concludes that if $m$ is integer
(half-integer),  $|\,\{m_i^a\}\,;j\,>$ and the highest
weight vector $|\,j,t>$ have the same (opposite)
statistics.

The algebra \uno\ is endowed with a linear  
anti-automorphism $\sigma$ defined as:
$$
\sigma(J_n^a)\,=\,J_{-n}^{-a}
\,\,\,\,\,\,\,\,\,\,\,\,\,\,
\sigma(j_n^{\alpha})\,=\,j_{-n}^{-\alpha}
\,\,\,\,\,\,\,\,\,\,\,\,\,\,
\sigma(k)\,=\,k\,\,.
\eqn\vuno
$$
Using $\sigma$ one can define an inner product for the
elements of $V^{(j,t)}$. In fact, given two arbitrary
elements $u$ and $v$ of $V^{(j,t)}$, they can be
represented in the form:
$$
u\,=\,u_-\,|\,j,t\,>
\,\,\,\,\,\,\,\,\,\,\,\,\,\,
v\,=\,v_-\,|\,j,t\,>
\,\,\,\,\,\,\,\,\,\,\,\,\,\,
u_-\,,\,v_-\,\,\in  U\,(\,{\cal A}_-\,)\,\,.
\eqn\vdos
$$
As one can naturally extend $\sigma$ to 
$ U\,(\,{\cal A})$, it makes sense to consider
$\sigma(u_-)$. The inner product $<\,u\,|\,v\,>$ is
obtained by acting with $\sigma(u_-)\,v_-$ on the highest
weight vector $|\,j,t\,>$ and taking the projection of
the result on the subspace $V^{(j,t)}_{0,0}$. If we define
$<\,j,t\,|\,\,j,t\,>=1$, the above definition reduces to:
$$
<\,u\,|\,v\,>\,\equiv\,<\,j,t\,|\,
\sigma(u_-)\,v_-\,|\,j,t\,>\,\,.
\eqn\vtres
$$
Eq. \vtres\ defines a bilinear form on $V^{(j,t)}$ , which
is usually called the contravariant (or Shapovalov) form.
Let us now introduce the states:
$$
e^{xJ_0^-\,+\,\theta j_0^-}\,\,\vert\,j,t>\,\,,
\eqn\vcuatro
$$
where $x$ is a complex number and $\theta$ a Grassmann
variable ($\theta^2=0$). On the states \vcuatro, the
zero-mode currents $J_0^a$ and $j_0^{\alpha}$, which
generate the finite ${\rm osp}(1\vert 2)$ superalgebra,
act as certain differential operators. Actually one can
prove that:
$$
\eqalign{
&J_0^a\,\,e^{xJ_0^-\,+\,\theta j_0^-}\,\,\vert\,j,t>\,=\,
D_j^{a}\,\,e^{xJ_0^-\,+\,\theta j_0^-}\,\,\vert\,j,t>\cr
&j_0^{\alpha}\,\,e^{xJ_0^-\,+\,\theta j_0^-}\,\,\vert\,j,t>\,=\,
d_j^{\alpha}\,\,e^{xJ_0^-\,+\,\theta j_0^-}\,
\,\vert\,j,t>\,\,,\cr}
\eqn\vcinco
$$
where $D^a_j$ and $d^{\alpha}_j$ are given by:
$$
\eqalign{
D^0_j\,=&\,-x\partial_x\,-\,{1\over 2}\,\theta\,
\partial_{\theta}\,+\,j\cr
D^+_j\,=&\,-x^2\partial_x\,+\,2jx\,-\,\theta x
\partial_{\theta}\cr
D^-_j\,=&\,\partial_x\cr
d^+_j\,=&\,x\partial_{\theta}\,+\,\theta x\partial_x\,
-\,2j\theta\cr
d^-_j\,=&\,\partial_{\theta}\,+\,\theta\partial_x\,\,.\cr}
\eqn\vseis
$$
Eq. \vcinco\ is a consequence of the highest weight
conditions \trece\ and of the following conjugation
formulas for the currents:
$$
\eqalign{
&e^{xJ_0^-\,+\,\theta j_0^-}\,J_n^0\,
e^{-xJ_0^-\,-\,\theta j_0^-}\,=\,J_n^0\,+\,xJ_n^-\,+\,
{\theta\over 2}\,j_n^{-}\cr
&e^{xJ_0^-\,+\,\theta j_0^-}\,J_n^+\,
e^{-xJ_0^-\,-\,\theta j_0^-}\,=\,J_n^+\,-\,2xJ_n^0\,-\,
x^2\,J_n^-\,+\,\theta\,(\,j_n^+\,-\,xj_n^-\,)\cr
&e^{xJ_0^-\,+\,\theta j_0^-}\,J_n^-\,
e^{-xJ_0^-\,-\,\theta j_0^-}\,=\,J_n^-\cr
&e^{xJ_0^-\,+\,\theta j_0^-}\,j_n^+\,
e^{-xJ_0^-\,-\,\theta j_0^-}\,=\,j_n^+\,-\,xj_n^-
\,+\,2\theta\,(\,J_n^0\,+\,xJ_n^-\,)\cr
&e^{xJ_0^-\,+\,\theta j_0^-}\,j_n^-\,
e^{-xJ_0^-\,-\,\theta j_0^-}\,=\,j_n^-\,
-\,2\theta\,J_n^-\,\,.
\cr}
\eqn\vsiete
$$
One can easily prove eq. \vsiete\ by using the defining
relations \uno\ of the algebra and the
Baker-Campbell-Hausdorff formula:
$$
e^A\,B\,e^{-A}\,=\,\sum_{p=0}^{\infty}\,\,
{1\over p!}\,\overbrace{[A,[A,[\cdots,
[A}^{p},B]\cdots]\,\,.
\eqn\vocho
$$

The primary fields are fundamental objects in any
Conformal Field Theory. The holomorphic part of one of
such fields depends on the coordinate $z$. If we have
some internal symmetry ${\cal G}$ in our theory, these
primary fields are associated to representations of 
${\cal G}$ and, in general, will have several components,
which keep track of the representation space of 
${\cal G}$. In our case, the internal symmetry is the 
${\rm osp}(1\vert 2)$ zero-mode superalgebra which, as we
have seen in eq. \vcinco, is represented by differential
operators in the isotopic variables $x$ and $\theta$. It
is thus natural to think that the primary fields should
depend also on these  variables. We shall denote by 
$\phi_j(z,x,\theta)$ the primary field corresponding to
the isospin $j$ representation.

The basic property that characterizes the primary fields
in Conformal Field Theory is the fact that their insertion
at the origin of coordinates $z=0$ creates a highest
weight state from the vacuum of the theory. As $L_{-1}$
is the translation operator in the variable $z$, the
change of the insertion point is equivalent to the action
of the operator $e^{zL_{-1}}$ on the $z=0$ state.
Moreover, the $x$ and $\theta$ dependence of the states
can be naturally introduced as in \vcuatro. All these
considerations lead us to define the action of 
$\phi_j(z,x,\theta)$ on the vacuum $\vert\,\Omega\,>$ of
the theory as:
$$
\phi_j(z,x,\theta)\,\vert\,\Omega\,>\,\equiv\,
e^{zL_{-1}\,+\,xJ_0^-\,+\,\theta j_0^-}\,
\vert\,j,t>\,\,.
\eqn\vnueve
$$
We shall assume that the vacuum state $\vert\,\Omega\,>$
is bosonic, which means that the Grassmann parity of 
$\phi_j(z,x,\theta)$ is equal to $p(j)$.

The commutators of the currents with the fields 
$\phi_j(z,x,\theta)$ is given by the standard expression:
$$
\eqalign{
&[\,J_n^a\,,\,\phi_j(z,x,\theta)\,]\,=\,z^n\,
D_j^{a}\,\phi_j(z,x,\theta)\cr
&[\,j_n^{\alpha}\,,\,\phi_j(z,x,\theta)\,]\,=\,z^n\,
d_j^{\alpha}\,\phi_j(z,x,\theta)\,\,,\cr}
\eqn\treinta
$$
where we have taken into account the implementation of
the ${\rm osp}(1\vert 2)$ finite superalgebra by the
differential operators \vseis. In order to write  
the second equation in \treinta, we have supposed that 
$\phi_j(z,x,\theta)$ has bosonic statistics. If this
were not the case, one should substitute the commutator
with $j_n^{\alpha}$ by the corresponding anticommutator. 
Moreover, the action of
the Virasoro modes $L_n$ on the primary fields 
$\phi_j(z,x,\theta)$ is determined by their conformal
weight $h_j$:
$$
[\,L_n\,,\,\phi_j(z,x,\theta)\,]\,=\,
[\,(n+1)\,h_j\,z^n\,+\,z^{n+1}\,\partial_z\,]\,
\phi_j(z,x,\theta)\,\,.
\eqn\tuno
$$

Let us now consider  the state obtained by acting with 
two primary fields, one of which is located at the origin
of coordinates, on the vacuum $\vert\,\Omega\,>$. This
state is:
$$
\phi_{j_2}(z,x,\theta)\,\,\phi_{j_1}(0,0,0)\,
\vert\,\Omega\,>\,\equiv\,\phi_{j_2}(z,x,\theta)\,
\vert\,j_1,t>\,\,.
\eqn\tdos
$$
The primary fields of the conformal field theory close
operator product algebras when they are multiplied. For
this reason, one can decompose the state \tdos\ as:
$$
\phi_{j_2}(z,x,\theta)\,\,\phi_{j_1}(0,0,0)\,
\vert\,\Omega\,>\,=\,\sum_{j_3}\,
\vert\,j_3,t,z,x,\theta>\,\,.
\eqn\ttres
$$
In eq. \ttres,  $\vert\,j_3,t,z,x,\theta>$ is a
coordinate-dependent state of $V^{(j_3,t)}$. When a state
of isospin $j_3$ is included in the right-hand side of eq.
\ttres\ we will say that the Verma module $V^{(j_3,t)}$
appears in the fusion of $V^{(j_1,t)}$ and $V^{(j_2,t)}$.
The doubly graded decomposition \dseis\ implies the
following expansion of $\vert\,j_3,t,z,x,\theta>$:
$$
\vert\,j_3,t,z,x,\theta>\,=\,\sum_{n,m}\,
\vert\,j_3,t,z,x,\theta,n,m>\,\,,
\eqn\tcuatro
$$
where the vectors appearing in the right-hand side of
\tcuatro\ have well-defined $L_0$ and $J_0^0$
eigenvalues. It is not difficult to obtain the coordinate
dependence of the states resulting from the fusion
\ttres. We shall verify in a moment that this dependence
is fixed by the covariance constraints satisfied by the
highest weight states. Indeed, the conditions \catorce\
for the vector $\vert\,j_1,t>$ read:
$$
\eqalign{
&(L_0\,-\,h_1)\,\vert\,j_1,t>\,=\,0\cr
&(J_0^0\,-\,j_1)\,\vert\,j_1,t>\,=\,0\,\,,\cr}
\eqn\tcinco
$$
where $h_i\equiv h_{j_i}$. Multiplying by 
$\phi_{j_2}(z,x,\theta)$ the two equations in \tcinco,
one gets:
$$
\eqalign{
&\phi_{j_2}(z,x,\theta)\,
(\,L_0\,-\,h_1\,)\,\vert\,j_1,t>\,=\,0\cr
&\phi_{j_2}(z,x,\theta)\,
(\,J_0^0\,-\,j_1\,)\,\vert\,j_1,t>\,=\,0\,\,.\cr}
\eqn\tseis
$$
Moreover, the commutation relations \treinta\ and \tuno\
imply that:
$$
\eqalign{
L_0\,\phi_{j_2}(z,x,\theta)\,=&\,
\phi_{j_2}(z,x,\theta)\,L_0\,+\,
(\,z\partial_{z}\,+\,h_2\,)\,\phi_{j_2}(z,x,\theta)\cr
J_0^0\,\phi_{j_2}(z,x,\theta)\,=&\,
\phi_{j_2}(z,x,\theta)\,J_0^0\,+\,
D_{j_2}^0\,\,\phi_{j_2}(z,x,\theta)\,\,.\cr}
\eqn\tsiete
$$
Using eq. \tsiete,  the constraints \tseis\ are converted
into:
$$
\eqalign{
(\,\,L_0\,-\,h_1\,-\,h_2\,-\,z\partial_z\,)\,
\phi_{j_2}(z,x,\theta)\,\vert\,j_1,t>\,=&\,0\cr
(\,J_0^0\,-\,j_1\,-\,D_{j_2}^0\,)\,
\phi_{j_2}(z,x,\theta)\,\vert\,j_1,t>\,=&\,0\,\,.\cr}
\eqn\tocho
$$
Substituting in \tocho\ the expansions \ttres\ and
\tcuatro, one can obtain constraints projected on a given
subspace $V^{(j_3,t)}_{(n,m)}$. As $L_0$ and $J_0^0$ act
diagonally on the vectors of $V^{(j_3,t)}_{(n,m)}$ with
eigenvalues $h_3+n$ and $j_3-m$ respectively, the
constraints on a given $(j_3, n, m)$ sector read:
$$
\eqalign{
(\,n\,+\,h_3\,-\,h_1\,-\,h_2\,-z\partial_z\,)\,
\vert\,j_3,t,z,x,\theta,n,m>\,=&\,0\cr
(\,j_3\,-\,j_1\,-\,j_2\,-\,m\,+\,x\partial_x\,+\,
{1\over 2}\,\theta\partial_{\theta}\,)
\vert\,j_3,t,z,x,\theta,n,m>\,=&\,0\,\,.\cr}
\eqn\tnueve
$$
Let us now assume the following general power dependence
for $\vert\,j_3,t,z,x,\theta,n,m>$:
$$
\vert\,j_3,t,z,x,\theta,n,m>\,=\,\theta^{\Delta_m}\,
z^A\,x^B\,\vert\,n,m>_{j_3}\,\,,
\eqn\cuarenta
$$
where $\vert\,n,m>_{j_3}$ is an element of
$V^{(j_3,t)}_{(n,m)}$ and, due to the Grassmann nature of
the variable $\theta$, $\Delta_m$  can only
take the values $0$ and $1$. The constraints \tnueve\
allow to determine the exponents in
\cuarenta. In fact, the substitution of the ansatz
\cuarenta\ in the $L_0$-condition (\ie\ the first equation
in \tnueve), yields the value of the $z$ exponent $A$: 
$$
A\,=\,n\,+\,h_3\,-\,h_1\,-\,h_2\,\,.
\eqn\cuno
$$
In the same way, the second equation in \tnueve\
implies that:
$$
B\,=\,m\,+\,j_1\,+\,j_2\,-\,j_3\,-\,{\Delta_m\over 2}\,\,.
\eqn\cdos
$$
It remains to determine $\Delta_m$. This can be done by
examining the Grassmann parity of the fusion relations.
Let us, first of all, introduce the  function 
$\epsilon(m)$, defined for an integer or half-integer
variable $m$ as follows:
$$
\epsilon(m)\,\equiv\,2(\,m\,-\,[m]\,)\,=\,
\cases{0&if $m\in\ZZ$\cr 
       1&if $m\in\ZZ+{1\over 2}\,\,.$\cr}
\eqn\ctres
$$

In eq. \ctres,  $[m]$ is the integer part of $m$ for any
$m\in \ZZ/2$, \ie\ $[m]=m$ when $m\in\ZZ$ and 
$[m]=m-{1\over 2}$ when $m\in\ZZ+{1\over 2}$. The
function $\epsilon(m)$ will be frequently used through
this paper. Let us list some of its (obvious) properties:
$$
\eqalign{
\epsilon(m\pm 1)\,=&\,\epsilon(m)
\,\,\,\,\,\,\,\,\,\,\,\,\,\,\,\,\,\,\,\,\,\,\,\,\,\,\,
\epsilon(m\pm {1\over 2})\,=\,1\,-\,\epsilon(m)\cr
(\,\epsilon(m)\,)^2\,=&\,\epsilon(m)
\,\,\,\,\,\,\,\,\,\,\,\,\,\,\,\,\,\,\,\,\,\,\,\,\,\,\,
\epsilon(m)\,\epsilon(m\pm {1\over 2})\,=\,0\,\,.\cr}
\eqn\ccuatro
$$
Coming back to the evaluation of $\Delta_m$, let us first
notice that the Grassmann parity of the vector
$\vert\,n,m>_{j_3}$ is $\epsilon (m)\,+\,p(j_3)\,\,$ 
${\rm mod}\,(2)$. Moreover,  it is also evident from our
previous equations that $\phi_{j_2}(z,x,\theta)\,
\vert\,j_1,t>$ and $\theta^{\Delta_m}\,\vert\,n,m>_{j_3}$
must have the same Grassmann parity. Therefore one must
have:
$$
p(j_1)\,+\,p(j_2)\,=\,\Delta_m\,+\epsilon (m)\,+\,p(j_3)
\,\,\,\,\,\,\,\,\,\,\,\,\,\,\,\,\,\,\,\,\,\,\,\,\,\,\,
{\rm mod}\,(2)\,\,.
\eqn\ccinco
$$
From eq. \ccinco\ one can easily find a closed
expression for $\Delta_m$ as a function of $m$ and  of the
parities
$p(j_i)$ of the highest weight vectors participating in
the fusion. One gets:
$$
\Delta_m\,=\,\epsilon(\,m\,+\,
{p(j_1)\,+\,p(j_2)\,-\,p(j_3)\over 2}\,)\,\,.
\eqn\cseis
$$
Therefore, the splitting \tcuatro\ can be written as:
$$
\vert\,j_3,t,z,x,\theta>\,=\,\sum_{n,m}\,
\theta^{\Delta_m}\,
z^{h_3-h_1-h_2+n}
\,x^{j_1+j_2-j_3+m-{\Delta_m\over 2}}
\,\vert\,n,m>_{j_3}\,\,.
\eqn\csiete
$$
Eqs. \cseis\ and \csiete\ will be of great importance 
in our analysis of the fusion of Verma modules. It is
interesting in what follows to obtain the range of the
numbers $n$ and $m$ in \csiete. First of all, it is clear
from its definition that $n$ is a non-negative integer.
Moreover, after analyzing the form of the basis elements
\dsiete, one easily concludes that $m$ is an integer or
half-integer number which is always greater or equal to
$-n$. Notice finally that the expansion \csiete\ is the
counterpart in our formalism of the operator product
expansion of the products of primary fields in CFT.  It is
worth to mention that, in general, the right-hand side of
eq. \csiete\ is not only singular in the variable $z$,
but also in the isotopic coordinate $x$.

\chapter{Decoupling of singular vectors}

For generic values of the isospins $j_i$, the fusion of
the Verma modules $V^{(j_1,t)}$ and $V^{(j_2,t)}$ in
$V^{(j_3,t)}$ is not restricted by any constraint.
However,  if one of the modules $V^{(j_i,t)}$ is
reducible, the situation changes completely. Indeed,  if
$V^{(j,t)}$ is reducible, it contains an unique maximal
submodule and we must formulate our Conformal Field
Theory in the module obtained by taking the quotient of 
$V^{(j,t)}$ by its maximal proper submodule. In this
quotient module, the vectors belonging to the submodule
vanish and, as we shall verify below, this condition
implies that the fusion of the corresponding primary
fields is not possible unless their isospins satisfy
non-trivial polynomial relations.

A Verma module $V^{(j,t)}$ is irreducible if and only if
it contains no singular vectors. These are vectors of 
$V^{(j,t)}$ which are annihilated by ${\cal A}_+$ and
have vanishing projection on $V^{(j,t)}_{0,0}$. It is
easy to prove that if $V^{(j,t)}$ has a singular vector,
the contravariant form on $V^{(j,t)}$ is degenerate.
Indeed, if $u$ is a singular vector and
$v=v_-\,|\,j\,,t\,>$ is an arbitrary vector of
$V^{(j,t)}$ ($v_-\in U({\cal A}_-)$), their inner product 
$<v|u>=<j,t|\sigma(v_-)u>$ vanishes since 
$\sigma (v_-)\in U({\cal A}_+)$. Therefore one can use
the determinant of the contravariant form in the
different subspaces  $V^{(j,t)}_{n,m}$  to locate the
singular vectors.  For bosonic affine algebras this
problem has been addressed in ref. 
\REF\KK{V. G. Kac and D. A. Kazhdan\journal\adm&34(79)97.}
[\KK]. The case of the 
${\rm osp}(1\vert 2)$ affine superalgebras has been
studied in refs. 
\REF\yudos{J. B. Fan and M. Yu, ``Modules over affine Lie
superalgebras", Academia Sinica preprint AS-ITP-93-14,
hep-th/9304122.} [\KW, \yudos]. Let us briefly review the 
${\rm osp}(1\vert 2)$ results. For a given value of $t$,
the singular vectors appear in Verma modules with highest
weight vectors whose isospins belong to a discrete set 
labelled by two integers $r$ and $s$. These isospins are
of the form:
$$
4\,j_{r,s}\,+\,1\,=\,r\,-\,st\,\,,
\eqn\cocho
$$
where $r+s$ is odd and either $r>0$ and $s\ge 0$ or 
$r<0$ and $s<0$. The $L_0$ and $J_0^0$ grades of the 
$V^{(j_{r,s},t)}$ subspace to which the singular vectors
belong are $n=rs/2$ and $m=r/2$ respectively.

Once one has determined under which conditions singular
vectors exist, one can try to find their explicit
expressions. For bosonic affine algebras these
expressions have been found by Malikov, Feigin and Fuks
(MFF) in ref. [\MFF]. These authors have found an equation
giving the singular vectors in terms of monomials
involving complex exponents of the generators. The
corresponding analysis for ${\rm osp}(1\vert 2)$ has been
performed in refs. [\KW, \yudos]. In general,  the singular
vector of the $V^{(j_{r,s},t)}$ ${\rm osp}(1\vert 2)$
module is given by:
$$
|\,\chi_{r,s}^{\pm}\,>\,=\,
F^{\pm}(r,s,t)\,|\,j_{r,s}\,,t\,>\,\,,
\eqn\cnueve
$$
where the $\pm$ index refers to the two possible signs of
$r$ and $F^{\pm}(r,s,t)$ is an element of 
$U({\cal A}_-)$. For  $r>0$ and $s\ge 0$, $F^{+}(r,s,t)$
is given by:
$$
\eqalign{
F^{+}(r,s,t)\,=&\,j_0^{-}\,
(\,J_0^{-}\,)^{{r-1+st\over 2}}\,\,
(\,J_{-1}^{+}\,)^{{r+(s-1)t\over 2}}\,\,
j_0^{-}\,(\,J_0^{-}\,)^{{r-1+(s-2)t\over
2}}\,\,\cdots\times\cr 
&\times\cdots\,\,
(\,J_{-1}^{+}\,)^{{r-(s-1)t\over 2}}\,\,
j_0^{-}\,(\,J_0^{-}\,)^{{r-1-st\over 2}}\,\,,\cr}
\eqn\cincuenta
$$
while, on the other hand, for $r<0$ and $s< 0$ the
corresponding operator $F^{-}(r,s,t)$ is:
$$
\eqalign{
F^{-}(r,s,t)\,=&\,(\,J_{-1}^{+}\,)^{-{r+(s+1)t\over 2}}\,\,
j_0^{-}\,(\,J_0^{-}\,)^{-{r+1+(s+2)t\over 2}}\,\,
(\,J_{-1}^{+}\,)^{-{r+(s+3)t\over 2}}\,\,\cdots\times\cr
&\times\cdots
j_0^{-}\,(\,J_0^{-}\,)^{-{r+1-(s+2)t\over 2}}\,\,
(\,J_{-1}^{+}\,)^{-{r-(s+1)t\over 2}}\,\,.\cr}
\eqn\ciuno
$$
It is far from obvious that the expressions \cincuenta\
and \ciuno\ define an element of $U({\cal A}_-)$. In
order to check this fact one must use some identities for
products of operators that involve general complex
powers. An example of such an identity is the following:
$$
A\,B^{\gamma}\,=\,\sum_{i=0}^{\infty}\,\,
{\gamma\choose i}\,B^{\gamma-i}\,
[\cdots[A,\overbrace{B],B],\cdots,B]}^{i}\,\,,
\eqn\cidos
$$
where $\gamma$ is not necessarily a non-negative integer.
In eq. \cidos,  the $i=0$ term should be understood as
$B^{\gamma}A$. Eq. \cidos\ can be regarded as the
analytical continuation of the case in which 
$\gamma\in\ZZ_+$. In this  case,  only a finite
number of terms contribute to the right-hand side of
\cidos\ and this identity is easily proved by induction.

Although we shall not reproduce here the proof [\KW,
\yudos] that 
$F^{\pm}(r,s,t)\in U({\cal A}_-)$, we shall give some
arguments in support of this result. First of all,  let us
notice that, as can be  checked by an elementary
calculation, the sum of the exponents of $J^+_{-1}$ in
\cincuenta\ and \ciuno\ is $rs/2$, which is always integer 
and non-negative since $r+s$ is odd. This property is
crucial if we want to end with a non-negative integer
power of 
$J^+_{-1}$ after applying \cidos\ to the operators
\cincuenta\ and \ciuno. In the same way,  one can verify
that the sum of the exponents of $J_0^-$ in 
$F^{\pm}(r,s,t)$ is $(r\mp 1)(s+1)/2$, which, again,  
always belongs to $\ZZ_+$. Finally,  it is interesting to
point out that there are $\pm (s+1)$ fermionic currents
$j_0^-$ in $F^{\pm}(r,s,t)$, which implies that the
operator $F^{\pm}(r,s,t)$ is bosonic (fermionic) if $r$ is
even (odd).

Let us now study the consequences, for the fusion of Verma
modules, of the  existence of singular vectors. Let us
assume that the isospin $j_1$ is of the form \cocho\
for some  $r=r_1$  and $s=s_1$, \ie\ that 
$j_1\,=\,j_{r_1,s_1}$. In the quotient module of
$V^{(j_1,t)}$ one should have:
$$
F^{\pm}(r_1,s_1,t)\,\,|\,j_1\,\,,t\,>\,=\,0\,\,.
\eqn\citres
$$
Let us see how eq. \citres\ restricts the fusion of
$V^{(j_1,t)}$ with another module $V^{(j_2,t)}$.
Multiplying eq. \citres\ by the primary field 
$\phi_{j_2}\,(z,x,\theta)$, one obviously gets:
$$
\phi_{j_2}\,(z,x,\theta)\,
F^{\pm}(r_1,s_1,t)\,\,|\,j_1\,,t\,>\,=\,0\,\,.
\eqn\cicuatro
$$
In order to derive from eq. \cicuatro\ a constraint for 
$\phi_{j_2}\,(z,x,\theta)\,|\,j_1\,,t\,>$, one should
exchange the order of $\phi_{j_2}\,(z,x,\theta)$ and 
$F^{\pm}(r_1,s_1,t)$ in this equation. As
$F^{\pm}(r_1,s_1,t)$ is an element of $ U({\cal A}_-)$,
this can be done by using eq. \treinta. However,  we are
interested in keeping the factorized form \cincuenta\ and
\ciuno\ of $F^{\pm}(r_1,s_1,t)$. Therefore we shall keep
the non-integer exponents in the expression of the
singular vectors. The commutation of
$\phi_{j_2}\,(z,x,\theta)$ with the $j_0^-$ terms in
\cincuenta\ and \ciuno\ is easy to calculate. In fact, if
the field $\phi_{j_2}\,(z,x,\theta)$ is bosonic, eq.
\treinta\ yields:
$$
\eqalign{
\phi_{j_2}\,(z,x,\theta)\,j_0^{-}\,=&\,
j_0^{-}\,\phi_{j_2}\,(z,x,\theta)\,+\,
[\,\phi_{j_2}\,(z,x,\theta)\,,j_0^{-}\,]\,=\,\cr
=&\,(\,j_0^{-}\,-\,d_{j_2}^{-}\,)\,
\phi_{j_2}\,(z,x,\theta)\,\,.\cr}
\eqn\cicinco
$$
When $p(j_2)=1$ a similar calculation, using the
anticommutator of $\phi_{j_2}\,(z,x,\theta)$ and $j_0^-$, 
can be performed. The final result differs from \cicinco\
in a global sign and, therefore, one can write in general:
$$
\phi_{j_2}\,(z,x,\theta)\,j_0^{-}\,=\,
(-1)^{p(j_2)}\,
(\,j_0^{-}\,-\,d_{j_2}^{-}\,)\,\phi_{j_2}\,(z,x,\theta)
\,\,.
\eqn\ciseis
$$

The commutation of $\phi_{j_2}\,(z,x,\theta)$ with the 
$J_0^-$ and $J_{-1}^+$ factors of $F^{\pm}(r_1,s_1,t)$ is
more delicate since these currents have non-integer
powers in \cincuenta\ and \ciuno. We shall use, to deal
with this case, the expression \cidos. Let us consider
the commutation of $J_0^-$ and
$\phi_{j_2}\,(z,x,\theta)$ first.  The iterated
commutators one has to compute in this case are: 
$$
[\cdots[\,\phi_{j_2}\,(z,x,\theta)\,,
\overbrace{J_0^{-}\,]\,
,J_0^{-}\,],\cdots,J_0^{-}\,]}^{i}\,=\,
[-D_{j_2}^{-}\,]^{i}\,\phi_{j_2}\,(z,x,\theta)\,\,,
\eqn\cisiete
$$
and, therefore, one has:
$$
\phi_{j_2}\,(z,x,\theta)\,(\,J_0^{-}\,)^{\gamma}\,=\,
\,\sum_{i=0}^{\infty}\,\,
{\gamma\choose i}\,(J_0^{-})^{\gamma-i}\,
[-D_{j_2}^{-}\,]^{i}\,\phi_{j_2}\,(z,x,\theta)\,\,.
\eqn\ciocho
$$
The right-hand side of \ciocho\ can be taken as the
expansion of $(\,J_0^{-}\,-\,D_{j_2}^{-}\,)^{\gamma}$.
Therefore we shall rewrite \ciocho\ as:
$$
\phi_{j_2}\,(z,x,\theta)\,(\,J_0^{-}\,)^{\gamma}\,=\,
(\,J_0^{-}\,-\,D_{j_2}^{-}\,)^{\gamma}\,
\phi_{j_2}\,(z,x,\theta)\,\,.
\eqn\cinueve
$$
The same procedure can be applied to $J_{-1}^{+}$, with
the result:
$$
\phi_{j_2}\,(z,x,\theta)\,(\,J_{-1}^{+}\,)^{\gamma}\,=\,
(\,J_{-1}^{+}\,-\,z^{-1}D_{j_2}^{+}\,)^{\gamma}\,
\phi_{j_2}\,(z,x,\theta)\,\,.
\eqn\sesenta
$$
It is clear that, with this procedure, one ends up with the
following exchange relation:
$$
\phi_{j_2}\,(z,x,\theta)\,\,F^{\pm}(r_1,s_1,t)\,=\,
\tilde F^{\pm}_{j_2}(r_1,s_1,t)\,
\phi_{j_2}\,(z,x,\theta)\,\,,
\eqn\suno
$$
where the operator $\tilde F^{\pm}_{j_2}(r_1,s_1,t)$ is
obtained by changing $j_{0}^{-}\rightarrow (-1)^{p(j_2)}
(\,j_{0}^{-}-d_{j_2}^{-}\,)$, 
$J_{0}^{-}\rightarrow J_{0}^{-}-D_{j_2}^{-}$ and
$J_{-1}^{+}\rightarrow J_{-1}^{+}-z^{-1}D_{j_2}^{+}$ in 
$F^{\pm}(r_1,s_1,t)$. Using eq. \suno\ in \cicuatro\ one
arrives at:
$$
\tilde F^{\pm}_{j_2}(r_1,s_1,t)\,\phi_{j_2}\,(z,x,\theta)
\,\,|\,j_1\,,t\,>\,=\,0\,\,.
\eqn\sdos
$$

Notice that the operators $J_{-n}^a\,-\,z^{-n}D_{j_2}^a$
and $j_{-n}^a\,-\,z^{-n}d_{j_2}^a$ close the same algebra
as the currents  $J_{-n}^a$ and $j_{-n}^a$ when 
$J_{-n}^a$ and $j_{-n}^a$ belong to ${\cal A}_-$. For
this reason $\tilde F^{\pm}_{j_2}(r_1,s_1,t)$ can be
arranged, by using the analytically continued commutation
relations \cidos, as a polynomial expression in the
operators $J_{-n}^a\,-\,z^{-n}D_{j_2}^a$
and $j_{-n}^a\,-\,z^{-n}d_{j_2}^a$ with exponents which
are non-negative integers. The proof of this fact is the
same that serves to demonstrate that 
$F^{\pm}(r_1,s_1,t)\in U({\cal A}_-)$. Therefore
the operator $\tilde F^{\pm}_{j_2}(r_1,s_1,t)$ is
unambiguously defined and it makes sense to project eq.
\sdos\ on the state $|\,j_3\,,t\,>$. After doing this
projection,  only the part of 
$\tilde F^{\pm}_{j_2}(r_1,s_1,t)$ containing the
derivatives survives and, eliminating some global factor,
one ends up with an equation of the type:
$$
\hat F^{\pm}_{j_2}(r_1,s_1,t)\,
<\,j_3,t|\phi_{j_2}\,(z,x,\theta)
\,\,|\,j_1\,,t\,>\,=\,0\,\,,
\eqn\stres
$$
where  $\hat F^{\pm}_{j_2}(r_1,s_1,t)$ is the differential
operator obtained from $F^{\pm}(r_1,s_1,t)$
by making the substitutions 
$j_{0}^{-}\rightarrow d_{j_2}^{-}$, 
$J_{0}^{-}\rightarrow D_{j_2}^{-}$ and 
$J_{-1}^{+}\rightarrow D_{j_2}^{+}$. The form of the
matrix element appearing in the left-hand side of eq.
\stres\ can be obtained from the formalism of fusion of
Verma modules developed in section 2. Indeed, in order to
get the value of 
$<\,j_3,t|\phi_{j_2}\,(z,x,\theta)\,\,|\,j_1\,,t\,>$, one
only needs to project the right-hand side of eq. \csiete\
on the $n=m=0$ sector. The result is:
$$
<\,j_3,t|\phi_{j_2}\,(z,x,\theta)
\,\,|\,j_1\,,t\,>\,=\,C_{123}\,\,
\theta^{\delta_j}\,\,z^{h_3-h_1-h_2}\,
x^{j_1+j_2-j_3-{\delta_j\over 2}}\,\,,
\eqn\scuatro
$$
where $C_{123}$ is a constant and $\delta_j$ is
$\Delta_m$ for $m=0$, \ie\ (see eq. \cseis):
$$
\delta_j\,\equiv\,\epsilon \Big(\,
{p(j_1)+p(j_2)-p(j_3)\over 2}\,\Big)\,\,.
\eqn\scinco
$$
Notice that $\delta_j$ takes the value zero if 
$p(j_3)\,=\,p(j_1)\,+\,p(j_2)$ $\,\,\,{\rm mod}\,\,(2)$ and
is equal to one when the parity of $|\,j_3\,,t\,>$ is not
the sum (modulo two) of those of $|\,j_1\,,t\,>$ and 
$|\,j_2\,,t\,>$.

The expressions we have for the operators 
$\hat F^{\pm}_{j_2}(r_1,s_1,t)$ involve powers of
derivatives with exponents that do not belong to $\ZZ_+$.
We have already argued that, after using the analytically
continued commutators, one gets  perfectly well-defined
differential operators $\hat F^{\pm}_{j_2}(r_1,s_1,t)$.
It is,  however,  more interesting for our purposes to keep
the non-integer powers  in
$\hat F^{\pm}_{j_2}(r_1,s_1,t)$  and to evaluate the
action of this operator on the matrix element \scuatro\
by means of the fractional calculus techniques
[\Fractional]. This method has been recently used in
connection with the free field representation of the
$sl(2)$ current algebra. Basically, one only needs to
evaluate fractional derivatives on functions that are
powers of the variables. In our case,  the general power
appearing in
\stres\ is of the form $\theta^{\gamma}\, x^{\lambda}$
with $\gamma\,=\,0,1$. The derivatives we need are:
$$
\eqalign{
(\,\partial_x\,)^n\,\theta^{\gamma}\,
x^{\lambda}\,=&\,
{\lambda !\over (\lambda-n)!}\,\theta^{\gamma}\,
x^{\lambda-n}\cr\cr
[\,D_{j_2}^{+}\,]^n\,\theta^{\gamma}\,
x^{\lambda}\,=&\,
{(2j_2-\lambda-\gamma)!\over
(2j_2-\lambda-\gamma-n)!}\,\theta^{\gamma}\,
x^{\lambda+n}\cr\cr
d_{j_2}^{-}\,[\,D_{j_2}^{-}\,]^n\,\theta^{\gamma}\,
x^{\lambda}\,=&\,
{\lambda !\over
(\lambda-n-1+\gamma)!}\,\theta^{1-\gamma}\,
x^{\lambda-n-1+\gamma}\,\,.\cr}
\eqn\sseis
$$
The results displayed in eq. \sseis\ can be proved by a
direct calculation when $n\in\ZZ_+$ and then they can be
analytically continued for complex $n$. The use of the
derivation rules \sseis\ allows to compute the left-hand
side of eq. \stres. In fact, one can prove that:
$$
\hat F^{\pm}_{j_2}(r_1,s_1,t)\,\theta^{\delta_j}\,
x^{j-{\delta_j\over 2}}\,=\,
f^{\pm}_{r_1,s_1}\,(t)\,\theta^{\epsilon(
{s_1+1+\delta_j\over 2})}\,\,
x^{j-{r_1\over 2}-{1\over 2}
\epsilon({s_1+1+\delta_j\over 2})}\,\,,
\eqn\ssiete
$$
where $f^{\pm}_{r_1,s_1}\,(t)$ is a numerical factor and
$j$ is the following combination of the isospins:
$$
j\,=\,j_1\,+\,j_2\,-j_3\,\,.
\eqn\socho
$$
If $V^{(j_3,t)}$ appears in the fusion of $V^{(j_1,t)}$
and $V^{(j_2,t)}$, the matrix element \scuatro\ must be
non-vanishing. This can only occur when the constant
$C_{123}$ is different from zero and, in this case, the
fulfillment of eq. \stres\ requires the vanishing of the
factors 
$f^{\pm}_{r_1,s_1}\,(t)$. The explicit form of these
factors can be obtained by collecting the factorials
resulting from the fractional derivations. For $r_1>0$ and
$s_1\ge 0$, one has:
$$
\eqalign{
f^{+}_{r_1,s_1}\,(t)\,=&\,\prod_{n=0}^{s_1}\,
{\Bigl[\,j\,+\,
{1\over 2}[\,nt-\epsilon({n\over 2})-(-1)^n\delta_j\,]
\,\Bigr]!\over
\Bigl[\,j\,+\,
{1\over 2}[\,(s_1-n)t-r_1-1+
\epsilon({n\over 2})+(-1)^n\delta_j\,]
\,\Bigr]!}\,\times\cr\cr
&\times\,\prod_{m=1}^{s_1}\,
{\Bigl[\,2j_2\,-\,j\,+
{1\over 2}[\,(m-1-s_1)t\,+\,r_1
-\epsilon({m\over 2})-(-1)^m\delta_j\,]
\,\Bigr]!\over
\Bigl[\,2j_2\,-\,j\,-
{1\over 2}[\,mt\,
+\epsilon({m\over 2})+(-1)^m\delta_j\,]
\,\Bigr]!}\,\,,\cr\cr}
\eqn\snueve
$$
while if $r_1<0$ and $s_1< 0$, a similar calculation
allows to get $f^{-}_{r_1,s_1}\,(t)$:
$$
\eqalign{
f^{-}_{r_1,s_1}\,(t)\,=&\,\prod_{n=0}^{-s_1-2}\,
{\Bigl[\,j\,-\,
{1\over 2}[\,r_1+\epsilon({n\over 2})
-(s_1+1+n)t+(-1)^n\delta_j\,]
\,\Bigr]!\over
\Bigl[\,j\,-\,
{1\over 2}[\,(n+1)t+
\epsilon({n+1\over 2})-(-1)^n\delta_j\,]
\,\Bigr]!}\,\times\cr\cr
&\times\,\prod_{m=0}^{-s_1-1}\,
{\Bigl[\,2j_2\,-\,j\,-
{1\over 2}[\,\epsilon({m\over 2})-mt
+(-1)^m\delta_j\,]\,\Bigr]!\over
\Bigl[\,2j_2\,-\,j\,+
{1\over 2}[\,r_1\,
-\epsilon({m\over 2})-(s_1+1+m)t
-(-1)^m\delta_j\,]
\,\Bigr]!}\,\,.\cr\cr}
\eqn\setenta
$$
The expressions \snueve\ and \setenta\ can be simplified
by dividing the factorials appearing in their numerators
by those of their denominators. In general,  if $A$ and
$B$ are complex numbers such that  $A-B\in \ZZ_+$, one
has:
$$
{A!\over B!}\,=\,\prod_{i=0}^{A-B-1}\,(A-i)\,\,.
\eqn\stuno
$$

It can be verified that the products appearing in
\snueve\ and \setenta\ can be rearranged in such a way
that eq. \stuno\ can be applied to all the quotients of
factorials in these equations. In fact,  one can prove that
the first product in the expression of
$f^{+}_{r_1,s_1}\,(t)$ can be written as:
$$
\prod^{r_1-1}_{\,\,}\prod^{s_1}_
{{\!\!\!\!\!\!\!\!\!\!\!\!\!\!\!\!\!\!n=0
\,\,\,\,\,\,\,\,\,\,m=0
\atop \!\!\!\!\!\!\!\!\!\!\!\!\!\!\!\!
n+m\in2\ZZ+\delta_j}}\,\,\,\,
\Bigl(\,j\,-\,{n\over 2}\,+{m\over 2}\,t\,\Bigr)\,\,,
\eqn\stdos
$$
whereas the second product in \snueve\ can be put as:
$$
\prod^{r_1}_{\,\,}\prod^{s_1}_
{{\!\!\!\!\!\!\!\!\!\!\!\!\!\!\!\!n=1
\,\,\,\,\,\,\,\,\,\,m=1
\atop \!\!\!\!\!\!\!\!\!\!\!\!\!\!\!\!
n+m\in2\ZZ+\delta_j}}\,\,\,\,
\Bigl(\,2j_2\,-\,j\,+\,{n\over 2}\,-{m\over 2}
\,t\,\Bigr)\,\,,
\eqn\sttres
$$
and, therefore, $f^{+}_{r_1,s_1}\,(t)$ is given by:
$$
f^{+}_{r_1,s_1}\,(t)\,=\,
\prod^{r_1-1}_{\,\,}\prod^{s_1}_
{{\!\!\!\!\!\!\!\!\!\!\!\!\!\!\!\!\!\!n=0
\,\,\,\,\,\,\,\,\,\,m=0
\atop \!\!\!\!\!\!\!\!\!\!\!\!\!\!\!\!
n+m\in2\ZZ+\delta_j}}\,\,\,\,
\Bigl(\,j_1\,+\,j_2\,-\,j_3\,
-\,{n\over 2}\,+\,{m\over 2}\,t\,\Bigr)
\,\,\,\,
\prod^{r_1}_{\,\,}\prod^{s_1}_
{{\!\!\!\!\!\!\!\!\!\!\!\!\!\!\!\!n=1
\,\,\,\,\,\,\,\,\,\,m=1
\atop \!\!\!\!\!\!\!\!\!\!\!\!\!\!\!\!
n+m\in2\ZZ+\delta_j}}\,\,\,\,
\Bigl(\,j_2\,-\,j_1\,+\,j_3\,
+\,{n\over 2}\,-\,{m\over 2}\,t\,\Bigr)\,\,.
\eqn\stcuatro
$$
Similarly, the expression \setenta\ for
$f^{-}_{r_1,s_1}\,(t)$ can be shown to be equivalent to:
$$
f^{-}_{r_1,s_1}\,(t)\,=\,
\prod^{\!\!-r_1}_{\,\,}\,\,\,\prod^{-s_1-1}_
{{\!\!\!\!\!\!\!\!\!\!\!\!\!\!\!\!\!\!\!\!\!\!n=1
\,\,\,\,\,\,\,\,\,\,\,\,\,m=1
\atop \!\!\!\!\!\!\!\!\!\!\!\!\!\!\!\!
n+m\in2\ZZ+\delta_j}}\,\,\,\,
\Bigl(\,j_1\,+\,j_2\,-\,j_3\,
+\,{n\over 2}\,-\,{m\over 2}\,t\,\Bigr)
\,\,\,\,
\prod^{\!\!-r_1-1}_{\,\,}\,\,\,\prod^{-s_1-1}_
{{\!\!\!\!\!\!\!\!\!\!\!\!\!\!\!\!\!\!\!\!\!\!n=0
\,\,\,\,\,\,\,\,\,\,\,\,\,\,m=0
\atop \!\!\!\!\!\!\!\!\!\!\!\!\!\!\!\!
n+m\in2\ZZ+\delta_j}}\,\,\,\,
\Bigl(\,j_2\,-\,j_1\,+\,j_3\,
-\,{n\over 2}\,+\,{m\over 2}\,t\,\Bigr)\,\,.
\eqn\stcinco
$$

For a given value of $t$, the vanishing of
$f^{\pm}_{r_1,s_1}\,(t)$ imposes non-trivial polynomial
conditions to the isospins $j_1$, $j_2$ and $j_3$. These
conditions must be required in order to have the
representation of isospin $j_3$ in the fusion of those
with isospins $j_1$ and $j_2$. Notice that these
conditions are induced by the reducibility of
$V^{(j_1,t)}$. One could similarly find the relations
that must be imposed when some of the other two modules
has singular vectors. As we shall see in  next
section, for some types of representations these
conditions are strong enough to determine the selection
rules of the operator algebra of the theory. Before
finishing this section,  it is interesting to point out the
great similarity between the functions \stcuatro\ and
\stcinco\ and those corresponding to the $sl(2)$ current
algebra. The main difference between these two cases is
the appearance in the ${\rm osp}(1\vert 2)$ result of the
parameter $\delta_j$, which takes into account the
relative Grassmann parity of the highest weight vectors
participating in the fusion. We shall analyze in  next
section the implications of this statistics dependence in
the fusion structure of the algebra.

\chapter{Fusion rules for admissible representations}

In this section we particularize our analysis to the
so-called admissible representations of 
${\rm osp}(1\vert 2)$ [\KW, \yudos]. These representations
occur for rational values of the parameter $t$. In fact, 
we shall  assume in this section that $t$ is given by:
$$
t\,=\,{p\over p\,'}\,\,,
\eqn\stseis
$$
where $p$ and $p\,'$ are coprime positive integers such that 
$p+p\,'$ is even and $p$ and ${p+p\,'\over 2}$ are relatively
prime [\yudos]. For this value of the level $t$,  the
representations with isospins given by eq. \cocho\ with
$r\not= 0\,\,\,$ ${\rm mod}\,\,(p)$ are completely
degenerate. The proper maximal submodule of 
$V^{(j_{r,s},t)}$,  in this case,  is generated by two
singular vectors which give rise to a double line
embedding diagram, very similar to the one appearing in
the minimal models of the (super)Virasoro algebra. The
admissible representations correspond to the case in
which we restrict $r$ and $s$ to take values in the grid 
$1\le r\le p-1\,\,\,$, $0\le s\le p\,'-1$. It was shown in
ref. [\KW] that the characters of these representations
form a representation of the modular group. In appendix A
we recall the calculation of these characters and study
their relation with the ones corresponding to the minimal
supersymmetric models. The results presented in this
appendix generalize the relation, discovered in ref.
\REF\Mukhi{S. Mukhi and S. Panda\journal\np&B338(90)263.}
[\Mukhi],  between the $sl(2)$ admissible representations
and the minimal Virasoro models.

In this section we are going to prove, using the singular
vector decoupling conditions found in section 3, that the
primary fields corresponding to the admissible
representations close a well-defined fusion algebra. Our
result is very similar to the one established in refs.
[\AY, \FM] for the $sl(2)$ current algebra. We will find
two types of fusion rules which cannot be satisfied
simultaneously. In order to derive this result,  let us
consider the fusion of two isospins $j_1$ and $j_2$ given
by:
$$
\eqalign{
&4j_1+1\,=\,r_1\,-\,s_1t
\,\,\,\,\,\,\,\,\,\,\,\,\,\,\,\,\,\,\,\,\,\,\,\,\,\,
4j_2+1\,=\,r_2\,-\,s_2t\cr
&1\le r_1, r_2\le p-1
\,\,\,\,\,\,\,\,\,\,\,\,\,\,\,\,\,
\,\,\,\,\,\,\,\,\,\,\,\,\,\,\,\,\,
0\le s_1, s_2\le p\,'-1\,\,.\cr}
\eqn\stsiete
$$

As $j_1=j_{r_1,s_1}$ is of the form \cocho\ with $r_1>0$
and $s_1\geq 0$, the Verma module $V^{(j_1,t)}$ will have
a singular vector of the type \cincuenta. Therefore, one
should impose the singular vector decoupling condition 
$f_{r_1,s_1}^{+}\,(t)=0$. Moreover, the isospin $j_1$
can also be written as  $j_{r_1-p, s_1-p\,'}$ and, when  
$r_1$ and $s_1$ belong to the grid of the admissible
representations (see eq. \stsiete), one has that 
$r_1-p<0$ and $s_1-p\,'<0$. Therefore the module
$V^{(j_1,t)}$ also possesses a singular vector of the
type \ciuno, whose decoupling condition requires the
vanishing of  $f_{r_1-p,s_1-p\,'}^{-}\,(t)$. In conclusion, 
one has the following two conditions: 
$$
f_{r_1,s_1}^{+}\,(t)\,=\,
f_{r_1-p,s_1-p\,'}^{-}\,(t)\,=\,0\,\,.
\eqn\stocho
$$
The explicit expression of $f_{r_1,s_1}^{+}\,(t)$ is
given in eq. \stcuatro, whereas, using eq. \stcinco, the
condition $f_{r_1-p,s_1-p\,'}^{-}\,(t)\,=\,0$ takes the
form:
$$
\prod^{\!\!p-r_1}_{\,\,}\,\,\,\prod^{p\,'-s_1-1}_
{{\!\!\!\!\!\!\!\!\!\!\!\!\!\!\!\!\!\!\!\!\!\!\!\!\!n=1
\,\,\,\,\,\,\,\,\,\,\,\,\,\,\,\,m=1
\atop \!\!\!\!\!\!\!\!\!\!\!\!\!\!\!\!\!\!\!\!\!\!\!\!
n+m\in2\ZZ+\delta_j}}\,\,\,\,
\Bigl(\,j_1\,+\,j_2\,-\,j_3\,
+\,{n\over 2}\,-\,{m\over 2}\,t\,\Bigr)
\,\,\,\,
\prod^{\!\!p-r_1-1}_{\,\,}\,\,\,\prod^{p\,'-s_1-1}_
{{\!\!\!\!\!\!\!\!\!\!\!\!\!\!\!
\!\!\!\!\!\!\!\!\!\!\!\!\!n=0
\,\,\,\,\,\,\,\,\,\,\,\,\,\,\,\,\,\,\,\,m=0
\atop \!\!\!\!\!\!\!\!\!\!\!\!\!\!\!\!\!\!\!\!\!
n+m\in2\ZZ+\delta_j}}\,\,\,\,
\Bigl(\,j_2\,-\,j_1\,+\,j_3\,
-\,{n\over 2}\,+\,{m\over 2}\,t\,\Bigr)\,=\,0\,\,.
\eqn\stnueve
$$
We shall prove below that eq. \stocho\ forces $j_3$ to
take values corresponding to admissible representations.
There are, however, two possibilities of satisfying
\stocho\ that we shall discuss separately in two
subsections.

\section{Fusion rule I}

Let us first  consider the equation 
$f_{r_1,s_1}^{+}\,(t)=0$ . As the  expression \stcuatro\
of $f_{r_1,s_1}^{+}\,(t)$  is completely factorized, it
follows that $f_{r_1,s_1}^{+}\,(t)$  vanishes if and only
if,  at least,  one of its factors is zero. In eq.
\stcuatro\ there are two different products. Suppose that
one of  the factors in the first of these products 
vanishes. If this occurs, the isospin $j_3$ can be
written as:
$$
j_3\,=\,j_1\,+j_2\,-{n\over 2}\,+\,{m\over 2}t
\,\,\,\,\,\,\,\,\,{\rm with}\,\,\,
0\le n\le r_1-1 \,\,\,\,\,\,\,\,\,
0\le m\le s_1\,\,.
\eqn\ochenta
$$
Similar considerations can be applied to the 
$f_{r_1-p,s_1-p\,'}^{-}\,(t)=0$ condition (see eq.
\stnueve). Suppose that one of the factors in the first
product of the left-hand side of \stnueve\ vanishes. If
this were the case, $j_3$ would be given by:
$$
j_3\,=\,j_1\,+j_2\,+{\bar n\over 2}\,-\,{\bar m\over 2}t
\,\,\,\,\,\,\,\,\,{\rm with}\,\,\,
1\le \bar n\le p-r_1 \,\,\,\,\,\,\,\,\,
1\le \bar m\le p\,'-s_1-1\,\,.
\eqn\ouno
$$
In eqs. \ochenta\ and \ouno,  $n$, $m$, $\bar n$ and
$\bar m$ are integers that can take values in the range
indicated in these equations. If eqs. \ochenta\ and
\ouno\ are simultaneously satisfied, after subtracting
them,  one gets:
$$
(m+\bar m)t\,=\,\bar n\,+\,n\,\,.
\eqn\odos
$$
It is easy to see that eq. \odos\ cannot be satisfied.
Indeed, from eqs. \ochenta\ and \ouno\ it follows that 
$1\le m+\bar m\le p\,'-1$ and thus $(m+\bar m)t\not\in
\ZZ$, in flagrant contradiction with the right-hand side
of \odos. We thus conclude that the first product in
\stnueve\ cannot vanish if eq. \ochenta\ holds. In order
to satisfy simultaneously eqs. \ochenta\ and \stnueve, 
the only remaining possibility is the cancellation of one
of the factors in the second product of
$f_{r_1-p,s_1-p\,'}^{-}\,(t)$. In this case,  $j_3$ would be:
$$
j_3\,=\,j_1\,-j_2\,+{n'\over 2}\,-\,{m'\over 2}t
\,\,\,\,\,\,\,\,\,{\rm with}\,\,\,
0\le n'\le p-r_1-1 \,\,\,\,\,\,\,\,\,
0\le m'\le p\,'-s_1-1\,\,.
\eqn\otres
$$
Subtracting, as before, the two expressions for $j_3$
(eqs. \ochenta\ and \otres) one gets the following
parametrization for $j_2$:
$$
4j_2+1\,=\,n+n'+1\,-\,(m+m')t\,\,.
\eqn\ocuatro
$$
Notice that, as $1\le n+n'+1\le p-1$ and $0\le m+m'\le
p\,'-1$, the numbers
$ n+n'+1$ and $m+m'$ are in the grid of the admissible
representations and, thus one gets the following
parametrization for $r_2$ and $s_2$:
$$
r_2\,=\,n+n'+1
\,\,\,\,\,\,\,\,\,\,\,\,\,\,
s_2\,=\,m+m'\,\,.
\eqn\ocinco
$$
Using these expressions for $r_2$ and $s_2$ in \ochenta\
and \otres, one obtains that $j_3$ can be written as:
$$
4j_3+1\,=\,r_3\,-\,s_3t\,\,,
\eqn\oseis
$$
where $r_3$ and $s_3$ can be written in terms of $r_2$
and $s_2$ as:
$$
\eqalign{
r_3\,=&\,r_1\,+\,r_2\,-2n\,-\,1\,=\,
r_1\,-\,r_2\,+2n'\,+\,1\cr
s_3\,=&\,s_1\,+\,s_2\,-2m\,=\,
s_1\,-\,s_2\,+2m'\,\,.\cr}
\eqn\osiete
$$
As a consistency check, notice that 
from \osiete\ it follows that $r_3+s_3$ is odd if 
$r_1+s_1\,\,,\,\,r_2+s_2 \in 2\ZZ+1$. Moreover, by
varying $n$, $m$, $n'$ and $m'$ within the range displayed
in eqs. \ochenta\ and \otres, one can obtain the range of
allowed values for $r_3$ and $s_3$. After an
straightforward calculation we get:
$$
\eqalign{
&|\,r_1\,-\,r_2\,|\,+\,1\,\le\,r_3\,\le\,
{\rm min}\,
(r_1\,+\,r_2\,-1\,,\,2p\,-\,r_1\,-\,r_2\,-\,1\,)\cr
&|\,s_1\,-\,s_2\,|\,\le\,s_3\,\le\,
{\rm min}\,
(s_1\,+\,s_2\,,\,2p\,'\,-\,s_1\,-\,s_2\,-\,2\,)\,\,.\cr}
\eqn\oocho
$$

It is interesting to point out that,  from our previous
equations,  it follows that $r_3$ and $s_3$ are always in
the range allowed to the admissible representations. To
establish this fact, let us notice that, eliminating $r_2$
and $s_2$ in \osiete\ by means of \ocinco, one gets that 
$r_3=r_1+n'-n$ and $s_3=s_1+m'-m$. If we freely vary $n$,
$m$, $n'$ and $m'$ in the ranges appearing in \ochenta\
and \otres\ (which is equivalent to consider different
values of $r_2$ and $s_2$ for fixed values of $r_1$ and
$s_1$), we get that $1\le r_3\le p-1$ and 
$0\le s_3\le p\,'-1$, which, as claimed, corresponds to the
grid of values of the admissible representations.

In order to completely identify  the representations
resulting from the fusion of $j_1$ and $j_2$, one should
determine the parity of their highest weight vectors. In
general, for an admissible representation with isospin
$j_{r,s}$, let us define the following quantity:
$$
\lambda_{r,s}\,=\,{r+s-1\over 4}\,=\,
j_{r,s}\,+\,{s\over 4}\,(1+t)\,\,.
\eqn\onueve
$$
Notice that, since $r+s$ is always odd,
$2\lambda_{r,s}\in \ZZ$. It turns out that the Grassmann
parity of the representations resulting from the fusion
is determined by the difference:
$$
\Delta\lambda\,=\,\lambda_{r_1,s_1}\,+\,
\lambda_{r_2,s_2}\,-\,\lambda_{r_3,s_3}\,\,.
\eqn\noventa
$$
Using eqs. \osiete\ and \onueve\ one can evaluate
$\Delta\lambda$ with the result:
$$
\Delta\lambda\,=\,{n+m\over 2}\,\,.
\eqn\nuno
$$
Recall (see eq. \stcuatro) that $n+m\in 2\ZZ+\delta_j$.
Therefore, it follows that if $\delta_j=0$ ($\delta_j=1$),
\ie\ if $p(j_3)\,=\,p(j_1)\,+\,p(j_2)\,\,\,\,{\rm mod}(2)$ 
($p(j_3)\,=\,p(j_1)\,+\,p(j_2)\,+\,1
\,\,\,\,{\rm mod}(2)$), then  $\Delta\lambda\in\ZZ$ 
($\Delta\lambda\in\ZZ\,+\,{1\over 2}$ ). Thus we can
write: 
$$
p(j_3)\,=\,p(j_1)\,+\,p(j_2)\,+\,2\Delta\lambda
\,\,\,\,\,\,\,\,\,\,\,\,\,\,\,\,
{\rm mod}(2)\,\,.
\eqn\ndos
$$

To finish this subsection, let us write the fusion rules
we have found in a more convenient form. We shall denote
by $[\,r\,,\,s\,]$ the admissible representation with
isospin $j_{r,s}$. With this notation,  it follows from
\oocho\ that the fusion rules can be written as:
$$
[\,r_1\,,\,s_1\,]\,\times\,[\, r_2\,,\,s_2\,]\,=
\!\!\!
\sum_{r_3=|r_1-r_2|+1} 
^{{\rm min}\,(\,r_1+r_2-1\,,\,2p-r_1-r_2-1\,)}
\,\,\,\,
\sum_{s_3=|s_1-s_2|} 
^{{\rm min}\,(\,s_1+s_2\,,\,2p\,'-s_1-s_2-2\,)}
\!\!\!\!\!\![\,r_3\,,\,s_3\,]\,.
\eqn\ntres
$$
One must keep in mind when using eq. \ntres\ that, as can
be seen from eq. \osiete, $r_3$ and $s_3$ jump in the sums
\ntres\ in steps of two units.

\section{Fusion rule II}

The possibility studied in section 4.1 is not the only way
to fulfill the decoupling conditions \stocho. Indeed, one
could satisfy the equation $f_{r_1,s_1}^{+}\,(t)=0$ by
requiring that one of the  factors appearing in the
second product of \stcuatro\ vanishes. Following the same
steps as in section 4.1, one can prove that this
requirement is incompatible with the vanishing of one of
the  factors of the second product in \stnueve.
Therefore,  one of the factors in the first product of 
$f_{r_1-p,s_1-p\,'}^{-}\,(t)$ must vanish and, in
conclusion, we must have:
$$
\eqalign{
j_3\,=&\,j_1\,-j_2\,-{n\over 2}\,+\,{m\over 2}t
\,\,\,\,\,\,\,\,\,{\rm with}\,\,\,
1\le n\le r_1 \,\,\,\,\,\,\,\,\,
1\le m\le s_1\cr
j_3\,=&\,j_1\,+j_2\,+{n'\over 2}\,-\,{m'\over 2}t
\,\,\,\,\,\,\,\,\,{\rm with}\,\,\,
1\le n'\le p-r_1 \,\,\,\,\,\,\,\,\,
1\le m'\le p\,'-s_1-1\,\,.\cr}
\eqn\ncuatro
$$

Subtracting the two equations in \ncuatro\ we can get
the value of $4j_2+1$. After adding $p-p\,'t=0$ to the
right-hand side of the resulting equation, one gets:
$$
4j_2+1\,=\,(\,p-n-n'+1\,)\,-\,(\,p\,'-m-m'\,)t\,\,.
\eqn\ncinco
$$
From eq. \ncinco,  one is tempted to identify $r_2$ and
$s_2$ with:
$$
r_2\,=\,p-n-n'+1
\,\,\,\,\,\,\,\,\,\,\,\,\,\,\,\,\,\,
s_2\,=\,p\,'-m-m'\,\,.
\eqn\nseis
$$
Varying $n$, $m$, $n'$ and $m'$ in the range written in
\ncuatro, one gets that $1\le r_2\le p-1$ and
$1\le s_2\le p\,'-2$. Notice that these values of $r_2$ and
$s_2$ belong to the grid of the admissible
representations. Moreover,  the range of $s_2$ reveals that
only for  $s_2>0$ and  $p\,'>2$ will this solution of eq.
\stocho\ take place. On the other hand, adding the two
expressions of $j_3$ in \ncuatro, we arrive at:
$$
4j_3+1\,=\,r_1+n'-n\,-\,(s_1+m'-m)t\,\,,
\eqn\nsiete
$$
which suggests the following identification of $r_3$ and
$s_3$:
$$
r_3\,=\,r_1+n'-n
\,\,\,\,\,\,\,\,\,\,\,\,\,\,\,\,\,\,
s_3\,=\,s_1+m'-m\,\,.
\eqn\nocho
$$
This identification can be confirmed by evaluating the
range of the possible values of $r_3$ and $s_3$, which,
after taking eq. \ncuatro\ into account, is  
$1\le r_3\le p-1$, $1\le s_3\le p\,'-2$. This result shows
that $s_3$ cannot be zero. Using eq. \nseis\ in \nocho\
one can eliminate one of the two integer indices in the
right-hand side of \nocho\ in favor of $r_2$ and $s_2$:
$$
\eqalign{
r_3\,=&\,p+r_1-r_2+1-2n\,=\,-p+r_1+r_2+2n'-1\cr
s_3\,=&\,p\,'+s_1-s_2-2m\,=\,-p\,'+s_1+s_2+2m'\,\,.\cr}
\eqn\nnueve
$$
Notice that, again, $r_3+s_3\in 2\ZZ+1$, as a consequence
of the fact that  $r_1+s_1$ and $r_2+s_2$ are odd 
and that $p+p\,'$ is even. It is now straightforward to get
the range of variation of $r_3$ and $s_3$ for fixed
$r_1$, $s_1$, $r_2$ and $s_2$. The result is:
$$
\eqalign{
|\,p-r_1-r_2\,|\,+\,1\,\le& r_3\le \,p-1-|\,r_1-r_2\,|\cr
|\,p\,'-s_1-s_2-1\,|\,+\,1\,\le &
s_3\le \,p\,'-2-|\,s_1-s_2\,|\,\,.\cr}
\eqn\cien
$$
Therefore, we can write the following fusion rule:
$$
[\,r_1\,,\,s_1\,]\,\times\,[\, r_2\,,\,s_2\,]\,=
\!\!\!
\sum_{r_3=|p-r_1-r_2|+1} 
^{ p-1-|\,r_1-r_2\,| }
\,\,\,\,
\sum_{s_3=  |\,p\,'-s_1-s_2-1\,|\,+\,1 } 
^{ \,p\,'-2-|\,s_1-s_2\,|  }
\!\!\!\!\!\![\,r_3\,,\,s_3\,]\,\,,
\eqn\ctuno
$$
where, as in eq. \ntres, $r_3$ and $s_3$ jump in steps of
two units.

Let us now determine the statistics of the 
$[\,r_3\,,\,s_3\,]$ representation. As in the fusion rule
of section 4.1, the relevant quantity to consider is 
$\Delta \lambda$ (defined as in eqs. \onueve\ and
\noventa). An elementary calculation shows that in this
case:
$$
\Delta \lambda\,=\,-{n'+m'\over 2}\,+\,{p+p\,'\over 4}\,\,,
\eqn\ctdos
$$
and, therefore, $p(j_3)$ is given by:
$$
p(j_3)\,=\,p(j_1)\,+\,p(j_2)\,+\,2\Delta\lambda
\,+\,{p+p\,'\over 2}
\,\,\,\,\,\,\,\,\,\,\,\,\,\,\,\,
{\rm mod}(2)\,\,.
\eqn\cttres
$$
Let us point out before finishing this section that the
conditions \ncuatro\ are incompatible with 
equations \ochenta\ and \otres. This means that, as 
anticipated above, both sets of fusion rules cannot be
satisfied simultaneously.

\chapter{The descent equations}

Let us continue elaborating the formalism for the fusion
of Verma modules that was introduced in section 2. Our
objective will be the computation of the vectors
$\vert\,n,m>_{j_3}$ appearing in the expansion \csiete.
We shall verify that, generically, the determination of the
vectors $\vert\,n,m>_{j_3}$ can be performed once the
action of the elements of ${\cal A}_{+}$ on them is
known. The equations encoding this action will be called
the descent equations, following the denomination
introduced in ref. [\Bauer] for the $sl(2)$ current
algebra. Our derivation of these equations starts with the
highest weight conditions for the vector $\vert\,j_1,t>$:
$$
J_p^a\,\vert\,j_1,t>\,=\,j_p^{\alpha}\,\vert\,j_1,t>\,=\,0
\,\,\,\,\,\,\,\,\,\,\,\,\,\,\,\,\,\,\,\,\,\,\,\,\,\,\,
\forall\,\,(\,J_p^a\,,\,j_p^{\alpha}\,)\,\in\,
{\cal A}_{+}\,\,.
\eqn\ctcuatro
$$
Multiplying eq. \ctcuatro\ by $\phi_{j_2}(z,x,\theta)$
and commuting this field with the currents, one gets:
$$
\eqalign{
(\,J_p^a\,-\,z^p\,D_{j_2}^a\,)\,
\phi_{j_2}(z,x,\theta)\,\vert\,j_1,t>\,=&\,0\cr
(\,j_p^{\alpha}\,-\,z^p\,d_{j_2}^{\alpha}\,)\,
\phi_{j_2}(z,x,\theta)\,\vert\,j_1,t>\,=&\,0\,\,.\cr}
\eqn\ctcinco
$$
Substituting in this equation the expansions \ttres\ and
\csiete, one can determine the result of applying the
currents of ${\cal A}_{+}$ to the states resulting from
the fusion of $V^{(j_1,t)}$ and $V^{(j_2,t)}$. Let us
detail this determination for the bosonic currents.
Introducing the decompositions \ttres\ and \csiete\ in
the first equation \ctcinco\ and projecting on a given
isospin $j_3$, one arrives at:
$$
\eqalign{
&\sum_{n,m}\,\theta^{\Delta_m}\,
z^{h_3-h_1-h_2+n}
\,x^{j_1+j_2-j_3+m-{\Delta_m\over 2}}
\,J_p^a\,\vert\,n,m>_{j_3}\,=\,\cr
&=\,\sum_{n,m}\,z^{h_3-h_1-h_2+n+p}\,
D_{j_2}^a\,(\,\theta^{\Delta_m}\,
x^{j_1+j_2-j_3+m-{\Delta_m\over 2}}\,)\,
\vert\,n,m>_{j_3}\,\,.\cr}
\eqn\ctseis
$$
In \ctseis\ the left-hand side contains
$J_p^a\,\vert\,n,m>_{j_3}\,$, while in the right-hand side
the derivative $D_{j_2}^a$ only acts on the $\theta$ and
$x$ variables. Using eq. \vseis\ and comparing the terms
in both sides of eq. \ctseis\ with the same powers of
$\theta$ and $x$, one can extract the value of 
$J_p^a\,\vert\,n,m>_{j_3}$. Let us express this result in
terms of the following combination of the isospins:
$$
i_{\pm}\,=\,-j_3\,+\,j_1\,\pm\,j_2\,\,.
\eqn\ctsiete
$$
With this definition, the descent equations for the
bosonic currents read:
$$
\eqalign{
J_p^{+}\,\vert\,n,m>_{j_3}\,=&\,
(\,-i_-+\,1\,-\,m\,
-\,{\Delta_m\over 2}\,)\,\vert\,n-p,m-1>_{j_3}
\,\,\,\,\,\,\,\,\,\,\,\,\,\,\,\,\,\,
(p\geq 0)\cr
J_p^{0}\,\vert\,n,m>_{j_3}\,=&\,
-(\,{i_+\,+\,i_-\over 2}\,+\,m\,)\,\vert\,n-p,m>_{j_3}
\,\,\,\,\,\,\,\,\,\,\,\,\,\,\,\,\,\,
\,\,\,\,\,\,\,\,\,\,\,\,\,\,\,\,\,\,\,\,\,\,\,\,
(p\geq 1)\cr
J_p^{-}\,\vert\,n,m>_{j_3}\,=&\,
(\,i_+\,+\,1\,+\,m\,
-\,{\Delta_m\over 2}\,)\,\vert\,n-p,m+1>_{j_3}
\,\,\,\,\,\,\,\,\,\,\,\,\,\,\,\,\,\,\,\,\,\,\,\,
(p\geq 1)\,\,.\cr\cr}
\eqn\ctocho
$$
For the fermionic currents of ${\cal A}_+$ one proceeds
similarly. One must be specially careful in this case with
the signs and with the powers of the Grassmann variable
$\theta$. The final result is: 
$$
\eqalign{
j_p^{+}\,\vert\,n,m>_{j_3}\,=&\,(-1)^{\Delta_m}\,\,
[\,1\,+\,(\,i_-\,+\,m\,-\,
{3\over 2}\,)\,\Delta_m\,]\,
\vert\,n-p,m-{1\over 2}>_{j_3}
\,\,\,\,\,\,\,\,\,\,\,\,\,\,\,\,\,\,
(p\geq 0)\cr
j_p^{-}\,\vert\,n,m>_{j_3}\,=&\,(-1)^{\Delta_m}\,\,
[\,1\,+\,(\,i_+\,+\,m\,-\,
{1\over 2}\,)\,\Delta_m\,]\,
\vert\,n-p,m+{1\over 2}>_{j_3}
\,\,\,\,\,\,\,\,\,\,\,\,\,\,\,\,\,\,
(p\geq 1)\,\,.\cr\cr}
\eqn\ctnueve
$$

Notice the remarkable fact that the coefficients
multiplying the vectors appearing in the right-hand side
of eqs. \ctocho\ and \ctnueve\ are independent of the
current mode $p$. On the other hand, as a check
of the correctness of eqs. \ctocho\ and \ctnueve, one can
easily verify that the matrix elements of the currents of
${\cal A}_+$ displayed in these equations are compatible
with the  (anti)commutation relations of the algebra (eq.
\uno).

Let us now see how the descent equations can be used to
determine the vectors ${|\,n,m\,>}_{j_3}$. In general,
these vectors belong to the subspace $V^{(j_3,t)}_{n,m}$.
Therefore, they can be represented as a linear combination
of the elements of a basis of $V^{(j_3,t)}_{n,m}$. One of
such a basis was described in section 2 (see eq.
\dsiete). In terms of the vectors \dsiete\ one can write: 
$$
{|\,n,m\,>}_{j_3}\,=\,\sum_{\{m_i^a\}}\,\,
C_{\{m_i^a\}}\,\,|\,\{m_i^a\}\,;j_3\,>\,\,,
\eqn\ctdiez
$$
where $C_{\{m_i^a\}}$ are some constants and only those
vectors with $L_0$ and $J_0^0$ grades $n$ and $m$
respectively enter the sum \ctdiez\ (recall that for a
given sequence $\{m_i^a\}$ the values of $n$ and $m$ are
given in eqs. \docho-\veinte). It is clear that the
determination of ${|\,n,m\,>}_{j_3}$ is equivalent to 
obtaining  the constants $C_{\{m_i^a\}}$. The latter
can be determined with the help of the inner product
defined in \vtres. Indeed, let us suppose that we
multiply both sides of \ctdiez\ by another basis vector 
$|\,\{{\overline m}_i^{\,a}\}\,;\,j_3\,>\,
\in\,V^{(j_3,t)}_{n,m}$. Doing this we would get:
$$
<\,\,
\{{\overline m}_i^{\,a}\}\,;\,j_3|\,
{\,n,m\,>}_{j_3}\,=\,
\sum_{\{m_i^a\}}\,\,
C_{\{m_i^a\}}\,\,
<\,\{{\overline m}_i^{\,a}\}\,;\,j_3|\,
\,\{m_i^a\}\,;j_3\,>\,\,.
\eqn\ctonce
$$
We shall regard eq. \ctonce\ as a linear system of
equations whose unknowns are the constants
$C_{\{m_i^a\}}$. The inner products appearing in the
right-hand side of eq. \ctonce\ can be computed from the
explicit expression of the vectors 
$|\,\,\{m_i^a\}\,;j_3\,>$ ( see eq. \dsiete). Moreover,
the products 
$<\,\{{\overline m}_i^{\,a}\}\,;\,j_3|\,
{\,n,m\,>}_{j_3}$ can be evaluated with the help of the
descent equations. Indeed, the anti-automorphism
$\sigma$,   appearing in  \vtres, 
transforms  the elements of ${\cal A}_-$ appearing in the
definition \dsiete\ in currents of ${\cal A}_+$, whose
action on ${|\,n,m\,>}_{j_3}$ is given by eqs. \ctocho\
and \ctnueve. An explicit calculation gives the
following result:
$$
\eqalign{
<\,\,
\{{\overline m}_i^{\,a}\}\,;\,j_3|\,
{\,n,m\,>}_{j_3}\,=&\,
\Bigl(\,m^+\,-\,{i_++i_-\over 2}\,\Bigr)^{m_0}\,\,\,\,
\prod_{i=1}^{\,[m^++{\Delta_{m^+}\over 2}]}\,\,
\Big(\,-i_+\,+\,m^+\,+\,{\Delta_{m^+}\over 2}\,-\,i\,
\Bigr)\,\times\cr
&\times\,\,\prod_{i=1}^{\,[m^-+{\Delta_{m}\over 2}]}\,\,
\Big(\,-i_-\,-\,m\,-\,{\Delta_{m}\over 2}\,+\,i\,
\Bigr)\,\,.\cr}
\eqn\ctdoce
$$

If the contravariant form at grades $(n,m)$ is
non-degenerate, the system of equations \ctonce\ can be
solved for $C_{\{m_i^a\}}$ and, therefore, the vector 
${|\,n,m\,>}_{j_3}$ can be determined. In particular, if
for a given value of $(n,m)$ all the products in the
left-hand side of eq. \ctonce\ are zero 
and the contravariant form in non-degenerate, it follows
from \ctonce\ that the vector ${|\,n,m\,>}_{j_3}$ must
vanish. We are now going to see  that this actually
happens for some particular values of $i_{\pm}$. In fact, 
we shall prove that when $i_-$ ($i_+$) is integer or
half-integer, those vectors ${|\,n,m\,>}_{j_3}$ with a 
grade  $m$ greater (smaller) that a certain value vanish.
This truncation of the descent equations will be very
important in what follows and, for this reason, we are
going to describe it in detail.

Let us first consider  the case in which $m>0$. Let us
split $m^-$ in the upper limit of the last product in
\ctdoce\ as $m^-\,=\,m\,+\,m^+$ (see eq. \veinte). In
general, if $A$ and $B$ are integers or half-integers, the
integer part of their sum $[A+B]$ can be related to $[A]$
and $[B]$ by means of the equation:
$$
[A+B]\,=\,[A]\,+\,[B]\,+\,\epsilon(A)\epsilon(B)\,\,.
\eqn\cttrece
$$
In particular, taking in \cttrece\
$A\,=\,m\,+\,{\Delta_m\over 2}$ and $B\,=\,m^+$, one has:
$$
[\,m^{-}\,+\,{\Delta_m\over 2}\,]\,=\,
[\,m\,+\,{\Delta_m\over 2}\,]\,+\,[m^+]\,+\,
\epsilon(m^+)\epsilon(\,m\,+\,{\Delta_m\over 2}\,)\,\,.
\eqn\ctcatorce
$$
Therefore, as $m^+\ge 0$, it is clear that there exists a
factor in 
$<\,\, \{{\overline m}_i^{\,a}\}\,;\,j_3|\,
{\,n,m\,>}_{j_3}$ equal to:
$$
\prod_{i=1}^{\,[m+{\Delta_{m}\over 2}]}\,\,
\Big(\,-i_-\,-\,m\,-\,{\Delta_{m}\over 2}\,+\,i\,
\Bigr)\,\,.
\eqn\ctquince
$$

In general,  a product of the form
$\prod_{i=1}^{N}\,\,(A+i)$ is zero if and only if the
first (last) factor $A+1$ ($A+N$) is non-positive
(non-negative) and the last factor, \ie\ $A+N$, is
integer. Applying this result to eq. \ctquince,  it
follows that the product \ctquince,  vanishes if the
following three conditions are satisfied:
$$
\eqalign{
m\,\ge\,-i_{-}\,+\,1\,-\,{\Delta_m\over 2}\,,
\,\,\,\,\,\,\,\,\,\,\,\,\,\,\,\,\,\,
i_{-}\,\le\,\,-\,
{1\over 2}\,\epsilon(\,m\,+\,{\Delta_m\over 2})\,,
\,\,\,\,\,\,\,\,\,\,\,\,\,\,\,\,\,\,
i_{-}\,\in\,\ZZ\,+\,
{1\over 2}\,\epsilon(\,m\,+\,{\Delta_m\over 2})\,\,.
\cr\cr}
\eqn\ctdseis
$$
On the other hand, from eq. \cseis\ it follows that: 
$$
\epsilon(\,m\,+\,{\Delta_m\over 2})\,=\,\delta_j\,\,.
\eqn\ctdsiete
$$
This means that the truncation conditions \ctdseis\ can
be simplified and put as:
$$
\left.
\eqalign{
m\,\ge&\,-i_{-}\,+\,1\,-\,{\Delta_m\over 2}\cr
i_{-}\,\le&\,\,-\,{\delta_j\over 2}\cr
i_{-}\,\in&\,\ZZ\,+\,{\delta_j\over 2} \cr} 
\right\}\;\;\;\,\,\,\,\,\,\Longrightarrow\,\,\,\,\,\,\,
|\,n,m\,>_{j_3}\,=\,0\,\,.
\eqn\ctdocho
$$
Eq. \ctdocho\ implies that when $i_-$ is a non-positive
integer or half-integer (depending on the value of
$\delta_j$, \ie\ on the parity of $|\,j_3,t\,>$) all the
vectors $|\,n,m\,>_{j_3}$ vanish when $m>0$ is large
enough. 

In the case $m<0$ one can similarly demonstrate the
existence of a truncation that eliminates vectors with
large absolute value of $m$. We could prove this by
studying eq. \ctdoce\ for $m<0$. It is however simpler to
introduce a new basis for $V^{(j,t)}$, constituted by 
the vectors:
$$
\overline{|\,\{m_i^a\}\,;j\,>}\,=\,
\prod_{i=1}^{+\infty}\,\Bigl(\,j_{-i}^+\,\Bigr)^{2m_i^+}\,\,
\prod_{i=1}^{+\infty}\,\Bigl(\,J_{-i}^0\,\Bigr)^{m_i^0}\,\,
\prod_{i=0}^{+\infty}\,\Bigl(\,j_{-i}^-\,\Bigr)^{2m_i^-}\,\,
|\,j,t>\,\,.
\eqn\ctdnueve
$$
The inner products of these basis vectors and 
$|\,{\,n,m\,>}_{j_3}$ can be easily computed by using the
descent equations. The result is:
$$
\eqalign{
\overline{<\,\,
\{{\overline m}_i^{\,a}\}\,;\,j_3}|\,
{\,n,m\,>}_{j_3}\,=&\,
\Bigl(\,-m^-\,-\,{i_++i_-\over 2}\,\Bigr)^{m_0}\,\,\,\,
\prod_{i=1}^{\,[m^-+{\Delta_{m^-}\over 2}]}\,\,
\Big(\,-i_-\,-\,m^-\,-\,{\Delta_{m^-}\over 2}\,+\,i\,
\Bigr)\,\times\cr
&\times\,\,\prod_{i=1}^{\,[m^++{\Delta_{m}\over 2}]}\,\,
\Big(\,-i_+\,-\,m\,+\,{\Delta_{m}\over 2}\,-\,i\,
\Bigr)\,\,.\cr}
\eqn\ctveinte
$$
If, as before, 
when $m<0$ we split the upper limit of the last factor in
\ctveinte\ as $m^++{\Delta_{m}\over 2}\,=\,
-m+{\Delta_{m}\over 2}\,+\,m^-$, we find that the scalar
product \ctveinte\ contains a factor:
$$
\prod_{i=1}^{\,[-m+{\Delta_{m}\over 2}]}\,\,
\Big(\,-i_+\,-\,m\,+\,{\Delta_{m}\over 2}\,-\,i\,
\Bigr)\,\,.
\eqn\ctvuno
$$
After analyzing the situations in  which this product
vanishes, one arrives at the following truncation
conditions for $m<0$:
$$
\left.
\eqalign{
m\,\le&\,-i_{+}\,-\,1\,+\,{\Delta_m\over 2}\cr
i_{+}\,\ge&\,\,{\delta_j\over 2}\cr
i_{+}\,\in&\,\ZZ\,+\,{\delta_j\over 2} \cr} 
\right\}\;\;\;\,\,\,\,\,\,\Longrightarrow\,\,\,\,\,\,\,
|\,n,m\,>_{j_3}\,=\,0\,\,.
\eqn\ctvdos
$$
Notice that in eq. \ctvdos\ $i_{+}$ is a non-negative
integer or half-integer (depending again on $\delta_j$).
It is also interesting to point out that, when the
conditions in \ctvdos\ are satisfied, there are no
singular terms in the variable $x$  in the expansion
\csiete.

We finish this section by recalling that the truncations
\ctdocho\ and \ctvdos\ are only valid for the grades $n$
and $m$ in which the contravariant form is
non-degenerate. Therefore, in order to apply these
equations, one must be sure that there are not singular
vectors in the corresponding subspace $V^{(j_3,t)}_{n,m}$.

\chapter{The Sugawara recursion relations and the
singular vectors}

The Sugawara expression of the energy-momentum tensor
(eq. \siete) can be used to obtain a set of recursion
relations among the vectors $|\,n,m\,>_{j_3}$. In some
cases, these relations, together with the truncation
conditions \ctdocho\ and \ctvdos, will allow us to write
a finite system of linear equations, whose resolution
provides a very efficient way of solving the descent
equations. On the other hand, as it was the case  for the
$sl(2)$ current algebra [\Bauer], the fusion formalism can
be used to obtain the explicit form of the singular vectors
of the 
${\rm osp}(1\vert 2)$ affine algebra. The basic tools in
the computation of singular vectors will be precisely the
truncation equations of section 5  and the Sugawara
recursion relations which we are now going to derive. 

Our starting point will be the expression of the Virasoro
generators $L_n$ in terms of the currents $J_n^a$ and 
$j_n^{\alpha}$. This expression is readily obtained by
substituting the mode expansions \seis\ in the Sugawara
equation \siete\ and by identifying the result with eq.
\ocho. One gets:
$$
L_n\,=\,{1\over t}\,\sum_{p=-\infty}^{+\infty}\,\,
:\,\Bigl[\,2J_{n-p}^0\,J_{p}^0\,+\,
J_{n-p}^+\,J_{p}^-\,+\,
J_{n-p}^-\,J_{p}^+\,-\,
{1\over 2}\,j_{n-p}^+\,j_{p}^-\,+\,
{1\over 2}\,j_{n-p}^-\,j_{p}^+\,\Bigr]:\,\,.
\eqn\ctvtres
$$
If, in particular, we put $n=0$ in \ctvtres, we obtain:
$$
\eqalign{
{t\over 2}\,L_0\,-\,J_0^0\,
(\,J_0^0\,+\,{1\over 2}\,)\,=&\,
\sum_{p=1}^{+\infty}\,\,\Bigl[\,2J_{-p}^0\,J_{p}^0\,+\,
J_{-p}^+\,J_{p}^-\,-\,{1\over 2}\,j_{-p}^+\,j_{p}^-\,
\Bigr]\,+\cr
&+\,\sum_{p=0}^{+\infty}\,\,\Bigl[\,J_{-p}^-\,J_{p}^+\,+\,
{1\over 2}\,j_{-p}^-\,j_{p}^+\,\Bigr]\,\,.\cr}
\eqn\ctvcuatro
$$
Let us apply both sides of eq. \ctvcuatro\ to the vector 
$|n,m>_{j_3}$. From the action of $L_0$ and $J_0^0$ on
$|n,m>_{j_3}$, 
$$
\eqalign{
L_0\,|n,m>_{j_3}\,=&\,(\,h_3\,+\,n\,)\,|n,m>_{j_3}\,=\,
\Bigl(\,{2j_3(j_3+{1\over 2})\over t}\,+\,n\,\Bigr)
\,|n,m>_{j_3}\cr
J_0^0\,|n,m>_{j_3}\,=&\,(\,j_3\,-\,m\,)\,|n,m>_{j_3}
\,\,,\cr}
\eqn\ctvcinco
$$
one finds the action of the left-hand side of \ctvcuatro:
$$
\Bigl[\,{t\over 2}\,L_0\,-\,J_0^0\,
(\,J_0^0\,+\,{1\over 2}\,)\,\Bigr]\,|n,m>_{j_3}\,=\,
\overline{|n,m>}_{j_3}\,\,,
\eqn\ctvseis
$$
where we have defined:
$$
\overline{|n,m>}_{j_3}\,\equiv\,
\Bigl[\,{t\over 2}\,n\,+\,m(2j_3\,-\,m\,+{1\over 2}\,)\,
\Bigr]\,|n,m>_{j_3}\,\,.
\eqn\ctvsiete
$$
On the other hand, as a consequence of the normal
ordering, in the right-hand side of eq. \ctvcuatro\ the
currents of ${\cal A}_+$ are to the right of those
belonging to ${\cal A}_-$. Therefore, we can use the
descent equations \ctocho\ and \ctnueve\ to evaluate the
action of the right-hand side of \ctvcuatro\ on
$|n,m>_{j_3}$. Hence, taking eq. \ctvseis\ into account,
we get:
$$
\eqalign{
\overline{|n,m>}_{j_3}\,=&\,
\Bigl(\,i_+\,+\,1\,+\,m\,-\,
{\Delta_m\over 2}\,\Bigr)\,
\sum_{p=1}^{n}\,J_{-p}^{+}\,|n-p,m+1>_{j_3}\,-\cr
-&\,(\,i_+\,+\,i_-+2m\,)\,
\sum_{p=1}^{n}\,J_{-p}^{0}\,|n-p,m>_{j_3}\,+\cr
+&\,\Bigl(\,-i_-+\,1\,-\,m\,-\,
{\Delta_m\over 2}\,\Bigr)\,
\sum_{p=0}^{n}\,J_{-p}^{-}\,|n-p,m-1>_{j_3}\,-\,\cr
-&\,{1\over 2}\,(-1)^{\Delta_m}\,\Bigl[\,
1\,+\,(\,i_+\,+\,m\,-\,{1\over 2}\,)\,
\Delta_m\,\Bigr]\,
\sum_{p=1}^{n}\,j_{-p}^{+}\,|n-p,m+{1\over 2}>_{j_3}\,+\cr
+&\,{1\over 2}\,(-1)^{\Delta_m}\,\Bigl[\,
1\,+\,(\,i_-\,+\,m\,-\,{3\over 2}\,)\,
\Delta_m\,\Bigr]\,
\sum_{p=0}^{n}\,j_{-p}^{-}\,|n-p,m-{1\over 2}>_{j_3}\,\,.
\cr\cr}
\eqn\ctvocho
$$
Taking different values of $n$ and $m$ in eq. \ctvocho, 
one obtains the announced Sugawara recursion relations.
In fact, eq. \ctvocho\ has a triangular structure with
respect to a partial ordering for the couples $(n,m)$
that, according to ref. [\Bauer], we define as follows.
Given two couples $(n,m)$ and $(n',m')$ we will say that 
$(n,m)\,\le\,(n',m')$ if and only if $n\,\le \,n'$ and 
$n+m\,\le \,n'+m'$. With respect to this ordering, all the
terms in the right-hand side of \ctvocho\ precede to the
one in the left-hand side. Notice that the first couple
in this ordering is $(n,m)\,=\,(0,0)$. Actually, for
$n=m=0$, eq. \ctvocho\ is a trivial identity of the type
$0=0$ (recall that $n\ge 0$ and  $m$ cannot be less
than $-n$). Therefore, it is  possible to solve eq.
\ctvocho\ iteratively, starting from the trivial $n=m=0$
equation and considering values of $(n,m)$ in increasing
order.

Let us now study a property of the fusion states that
will allow us to use them in the computation of the
singular vectors. First of all, let us rewrite the descent
equations \ctocho\ and \ctnueve\ in the more condensed
form:
$$
\eqalign{
J_p^a\,\vert\,n,m>_{j_3}\,=&\,\mu_j^a(m)\,\,
\vert\,n-p,m-a>_{j_3}\cr
j_p^{\alpha}\,\vert\,n,m>_{j_3}\,=&\,\rho_j^{\alpha}(m)\,\,
\vert\,n-p,m-{\alpha\over 2}\,>_{j_3}
\,\,\,\,\,\,\,\,\,\,\,\,\,\,\,\,\,\,\,\,\,\,\,
\forall\,\,(\,J_p^a\,,\,j_p^{\alpha}\,)\,\in\,
{\cal A}_{+}\,\,,\cr\cr}
\eqn\ctvnueve
$$
where the coefficients $\mu_j^a(m)$ and
$\rho_j^{\alpha}(m)$ can be read from \ctocho\ and
\ctnueve. The vectors $\overline{\vert\,n,m>}_{j_3}$
satisfy:
$$
\eqalign{
&J_p^a\,\overline{\vert\,n,m>}_{j_3}\,=\,
[\,{t\over 2}\,n\,+\,m\,(2j_3\,-\,m\,+\,{1\over 2}\,)\,]
\,\,\mu_j^a(m)\,\,\vert\,n-p,m-a>_{j_3}\cr
&j_p^{\alpha}\,\overline{\vert\,n,m>}_{j_3}\,=\,
[\,{t\over 2}\,n\,+\,m\,(2j_3\,-\,m\,+\,{1\over 2}\,)\,]
\,\,\rho_j^{\alpha}(m)\,\,
\vert\,n-p,m-{\alpha\over 2}\,>_{j_3}\cr
&\forall\,\,(\,J_p^a\,,\,j_p^{\alpha}\,)\,\in\,
{\cal A}_{+}\,\,.\cr}
\eqn\cttreinta
$$
The proof of this equation is immediate if we use the
relation between the $\vert\,n,m>_{j_3}$ and 
$\overline{\vert\,n,m>}_{j_3}$ vectors (eq. \ctvsiete).
It turns out, however, that it can be proved acting with
the currents of ${\cal A}_{+}$ on the right-hand side of
eq. \ctvocho. In order to verify this statement one must
use the (anti)commutators of the algebra, together with
the descent equations \ctocho\ and \ctnueve. In fact, to
prove \cttreinta\ it is enough to check the cases
$J_p^a\,=\,J_1^-$ and $j_p^{\alpha}\,=\,j_0^{+}$, since
the result for the other currents can be proved by using
the algebra relations \uno. This implies that 
$\overline{\vert\,n,m>}_{j_3}$ would satisfy \cttreinta\
if we had defined it by means of eq. \ctvocho\ instead of
using eq. \ctvsiete. Another interesting observation is
that the prefactor 
${t\over 2}\,n\,+\,m\,(2j_3\,-\,m\,+\,{1\over 2}\,)$
appearing in the right-hand side of eq. \cttreinta\
vanishes  for $j_3=j_{r,s}$, $n={rs\over 2}$ and
$m\,=\,{r\over 2}$, where $r$ and $s$ are integers such
that $rs\ge 0$, $r+s\in\ZZ$ and $r\not= 0$ (this last
condition eliminates the trivial solution $n=m=0$).
Notice that these are precisely the isospins and grades
where the singular vectors are located. Therefore, as the
vector 
$\overline{\vert\,rs/ 2 \,,\,r/ 2\,>}_{j_{r,s}}$
satisfies:
$$
\eqalign{
J_p^a\,\overline{\vert\,rs/ 2
\,,\,r/ 2\,>}_{j_{r,s}}\,=\,
j_p^a\,\overline{\vert\,rs/ 2
\,,\,r/2\,>}_{j_{r,s}}\,=\,0\,,
\,\,\,\,\,\,\,\,\,\,\,\,\,\,\,\,\,\,\,\,\,\,\,
\forall\,\,(\,J_p^a\,,\,j_p^{\alpha}\,)\,\in\,
{\cal A}_{+}\,\,,\cr\cr}
\eqn\cttuno
$$
one is tempted to identify this vector with the
corresponding singular vector $\vert\,\chi_{r,s}\,>$:
$$
\vert\,\chi_{r,s}\,>\,\sim\,\overline{\vert\,rs/ 2
\,,\,r/ 2\,>}_{j_{r,s}}\,\,. 
\eqn\cttdos
$$
Obviously, the identification $\cttdos$  only makes sense
when the vector  $\overline{\vert\,rs/ 2\,,\,r/
2\,>}_{j_{r,s}}$ is not identically zero. In order to
have a non-trivial result we should find some criteria to
discard a priori the 
$\overline{\vert\,rs/ 2\,,\,r/ 2\,>}_{j_{r,s}}=0\,\,$
solution of the descent equations. In fact, for  general
values of $i_\pm$, it   is easy to
convince oneself that the recursion relations do not give
a result in which  $\overline{\vert\,rs/ 2\,,\,r/
2\,>}_{j_{r,s}}$  is identically zero.  For this reason, 
it is clear that $i_\pm$ must satisfy some non-trivial
conditions in order to get a vanishing result
for the vector \cttdos. The crucial observation [\Bauer] to
determine these conditions is that singular vectors do
vanish in the quotient of the Verma module by its maximal
proper submodule. The singular vector decoupling
conditions studied in section 3 are precisely the
requirements one has to impose to pass from the Verma
module to the corresponding quotient. Thus we expect that
these conditions are precisely the ones that $i_\pm$ must
satisfy in order to get 
$\overline{\vert\,rs/ 2\,,\,r/ 2\,>}_{j_{r,s}}=0\,\,$ as
the solution of the recursion relations \ctvocho. It is
thus clear that, in order to obtain a non-trivial result
for the singular vectors, one must be sure that the
decoupling conditions are not satisfied. Notice that eq.
\cttuno\ is satisfied independently of the decoupling
conditions and, therefore, even when the latter are not
satisfied, eq. \cttuno\ still holds.

Let us show how our previous considerations can be
applied in practice to the determination of the singular
vectors corresponding to the isopins $j_3=j_{r,s}$ with
$r>0$ and $s\ge 0$. The decoupling conditions of these
singular vectors in the module $V^{(j_3,t)}$ take the
form:
$$
g^{+}_{r,s}\,(i_+,i_-)\,=\,0\,\,,
\eqn\ctttres
$$
where the function $g^{+}_{r,s}\,(i_+,i_-)$ is given by:
$$
g^{+}_{r,s}\,(i_+,i_-)\,=\,
\prod^{r-1}_{\,\,}\prod^{s}_
{{\!\!\!\!\!\!\!\!\!\!\!\!\!\!\!\!\!\!i=0
\,\,\,\,\,\,\,\,\,\,l=0
\atop \!\!\!\!\!\!\!\!\!\!\!\!\!\!\!\!
i+l\in2\ZZ+\delta_j}}\,\,\,\,
\Bigl(\,-i_-\,
-\,{i\over 2}\,+\,{l\over 2}\,t\,\Bigr)
\,\,\,\,
\prod^{r}_{\,\,}\prod^{s}_
{{\!\!\!\!\!\!\!\!\!\!\!\!\!\!\!\!i=1
\,\,\,\,\,\,\,\,\,\,l=1
\atop \!\!\!\!\!\!\!\!\!\!\!\!\!\!\!\!
i+l\in2\ZZ+\delta_j}}\,\,\,\,
\Bigl(\,i_+\,
+\,{i\over 2}\,-\,{l\over 2}\,t\,\Bigr)\,\,.
\eqn\cttcuatro
$$
The form \cttcuatro\ of $g^{+}_{r,s}\,(i_+,i_-)$ can be
obtained from the function $f^{+}_{r_1,s_1}\,(t)$ written
in eq. \stcuatro\ by exchanging $j_1\leftrightarrow j_3$
and substituting $r_1$ and $s_1$ by $r$ and $s$
respectively.

Let us now discuss the election of the isospins $j_1$ and
$j_2$, \ie\ of $i_{\pm}$. In principle,  $j_1$ and $j_2$
can be arbitrarily chosen. However, as we have just
argued, we must require the condition 
$g^{+}_{r,s}\,(i_+,i_-)\not= 0$ in order to avoid having
a trivial result. Moreover, we can make use of the
truncation conditions of section 5 in order to deal with
the minimum number of intermediate vectors in the
recursion relation, which will make the singular vector
determination procedure more efficient. The truncation
conditions of eqs. \ctdocho\ and \ctvdos\ determine the
highest and lowest values of the $J_0^0$ grade $m$. Since
the singular vector is located at $m=r/2$, we shall
require that:
$$
\vert\,n\,,\,m\,>_{j_{r,s}}\,=\,0
\,\,\,\,\,\,\,\,\,\,\,\,\,\,\,\,\,
{\rm for}\,\,m\,\ge\,{r+1\over 2}\,\,.
\eqn\cttcinco
$$
Notice that for $m\,\ge\,{r+1\over 2}\,\,$ the
contravariant form is non-degenerate in 
$V^{(j_{r,s},t)}_{n,m}$, which ensures the validity of
the implication written in eq. \ctdocho. Actually, 
in view of this equation, one must have:
$$
{r+1\over 2}\,=\,-i_-\,+\,1\,-\,{1\over 2}\,
\Delta_{{r+1\over 2}}\,\,.
\eqn\cttseis
$$
Eq. \cttseis\ can be solved for $i_-$ for different
values, modulo two, of $r$ and $\delta_j$. One can easily
check that the result is 
$i_-\,=\,-{r\over 2}$ if $r+\delta_j\,=\,0\,\,\,$ 
${\rm mod}\,(2)$  and
$i_-\,=\,-{r-1\over 2}$ if $r+\delta_j\,=\,1\,\,$ 
${\rm mod}\,(2)$, or in a more compact form:
$$
i_-\,=\,-{r\over 2}\,\epsilon
\Bigl(\,{r+\delta_j+1\over 2}\,\Bigr)\,-\,
{r-1\over 2}\,\epsilon
\Bigl(\,{r+\delta_j\over 2}\,\Bigr)\,\,.
\eqn\cttsiete
$$
Of these two solutions, one can discard one of them by
looking at the value of the function 
$g^{+}_{r,s}\,(i_+,i_-)$. Indeed, it can be verified by
direct substitution that:
$$
g^{+}_{r,s}\,(\,i_+, -{r-1\over 2}\,)\,=\,0
\,\,\,\,\,\,\,\,\,\,\,\,\,\,\,\,\,
{\rm if}\,\,\,r+\delta_j\,=\,1 
\,\,\,{\rm mod}(2)\,\,,
\eqn\cttocho
$$
and, therefore, 
in order to have a non-trivial singular vector we
shall take $r+\delta_j\,=\,0$ mod(2), \ie:
$$
\delta_j\,=\,\epsilon(\,{r\over 2}\,)\,\,.
\eqn\cttnueve
$$
Eq. \cttnueve\ fixes the value of the parameter
$\delta_j$, which determines the relative Grassmann
parity of the modules involved in the fusion. For the
value \cttnueve\ of $\delta_j$, eq. \cttsiete\ gives the
value we must take for $i_-$, namely:
$$
i_-\,=\,-{r\over 2}\,\,.
\eqn\ctcuarenta
$$

Let us now choose $i_+$. It can be verified by inspection
that, for a generic value of $t$, the factor of 
$g^{+}_{r,s}\,(i_+,i_-)$ depending on $i_+$ never
vanishes. Therefore, we have in this case a larger freedom
in the election of $i_+$. Notice that, according to eq.
\ctvdos, $i_+$ determines the lower value in the range of
variation of $m$. We shall choose the smallest possible
value of $i_+$ which, according to eq. \ctvdos, gives
rise to  a value of $m$ that reduces
maximally the range of values that $m$ can take.
Therefore, taking eq. \ctvdos\ into account, we put:
$$
i_+\,=\,{\delta_j\over 2}\,=\,{1\over 2}\,
\epsilon(\,{r\over 2}\,)\,\,.
\eqn\ctcuno
$$
For this value of $i_+$ the vectors  
$|n, m>_{j_{r,s}}$  are zero if $m\,<\,-\delta_j/2$ and 
thus, in order to obtain the expression of the singular
vector for the isospin $j_{r,s}$, which we shall denote
by $|\,\lambda_{r,s}\,>$, we must vary  $n$ and $m$ in the
recursion relations  within the intervals:
$$
0\,\le\,n\,\le\,{rs\over 2}
\,\,\,\,\,\,\,\,\,\,\,\,\,\,\,\,\,\,
-{1\over 2}\,\epsilon(\,{r\over 2}\,)\,
\le\,m\,\le\,{r\over 2}\,\,.
\eqn\ctcdos
$$

We have now all the ingredients needed for the
computation of the singular vectors. In fact, as for the
value of $\delta_j$ written in eq. \cttnueve\ 
$\Delta_{r\over 2}\,=\,0$, one can write:
$$
\eqalign{
|\,\lambda_{r,s}\,>\,=&\,-{1\over 2}\,\,
(\,r\,+\,\epsilon(\,{r\over 2}\,)\,)\,
\sum_{p=1}^{{rs\over 2}}\,\,
J_{-p}^{0}\,|\,{rs\over 2}-p\,,\,{r\over 2}\,>_{j_{r,s}}
\,+\cr
&+\,\sum_{p=0}^{{rs\over 2}}\,\,
J_{-p}^{-}\, |\,{rs\over 2}-p\,,\, 
{r\over 2}-1\,>_{j_{r,s}}\, +\,{1\over 2}
\,\,\sum_{p=0}^{{rs\over 2}}\,\,j_{-p}^{-}\,
|\,{rs\over 2}-p\,,\, {r-1\over
2}\,>_{j_{r,s}}\,\,,\cr\cr}
\eqn\ctctres
$$
where we have used the values of $i_{\pm}$ of eqs.
\ctcuarenta\ and \ctcuno. The vectors appearing in the
right-hand side of eq. \ctctres\ are computed from the
recursion relations \ctvocho\ using the  values of
$i_{\pm}$ and  $\delta_j$ previously determined . In  
appendix B we present the detailed computation of the
vectors $|\,\lambda_{1,2}\,>$ and $|\,\lambda_{2,1}\,>$.
For more general values of $r$ and $s$ the calculation,
although  more involved, follows the same lines.

\chapter{The Knizhnik-Zamolodchikov equation}

The Sugawara recursion relations obtained in the previous
section can be alternatively derived from the existence
of a mixed Virasoro-Kac-Moody singular vector. This
vector is the same that gives rise to the well-known 
Knizhnik-Zamolodchikov equation 
\REF\KZ{V. G. Knizhnik and A. B.
Zamolodchikov\journal\np&B247(84)83.}
[\KZ],  which plays a
fundamental r\^ole in the determination of the correlation
functions of the theory. In this section we shall present
this alternative derivation of the descent equations
following the same method used in ref. [\Bauer] for the
$sl(2)$ algebra. 

It follows from the $L_{-1}$ expression (\ie\ from eq.
\ctvtres\ with $n=-1$) and from the highest weight
conditions \trece\ for the vector $|\,j_2\,,\,t\,>$ that:
$$
\Bigl[\,{t\over 2}\,L_{-1}\,-\,J_{-1}^+\,J_{0}^-\,-\,2
J_{-1}^0\,J_0^0\,+\,{1\over 2}\,j_{-1}^+\,j_0^-\,
\Bigr]\,|\,j_2\,,\,t\,>\,=\,0\,\,.
\eqn\ctccuatro
$$
As $J_0^0\,|\,j_2\,,\,t\,>\,=\,j_2\,|\,j_2\,,\,t\,>$, one
can obtain from \ctccuatro\ the following equation:
$$
e^{zL_{-1}\,+\,xJ_0^{-}\,+\,\theta j_{0}^-}\,
\phi_{j_1}\,(-z,-x,-\theta)\,\,
\Bigl[\,{t\over 2}\,L_{-1}\,-\,J_{-1}^+\,J_{0}^-\,-\,2
j_2\,J_{-1}^0\,+\,{1\over 2}\,j_{-1}^+\,j_0^-\,
\Bigr]\,|\,j_2\,,\,t\,>\,=\,0\,\,.
\eqn\ctccinco
$$
In the derivation of eq. \ctccinco\ from \ctccuatro\ we
have multiplied the latter by 
$e^{zL_{-1}\,+\,xJ_0^{-}\,+\,\theta j_{0}^-}\,
\phi_{j_1}\,(-z,-x,-\theta)$. Moreover, commuting 
$\phi_{j_1}\,(-z,-x,-\theta)$ with the currents in the
left-hand side of eq. \ctccinco, one finds:
$$
e^{zL_{-1}\,+\,xJ_0^{-}\,+\,\theta j_{0}^-}\,
G_{j_1, j_2 }(\,z,x,\theta\,)\,
\phi_{j_1}\,(-z,-x,-\theta)\,
\,|\,j_2\,,\,t\,>\,=\,0\,\,,
\eqn\ctcseis
$$
where $G_{j_1, j_2 }(\,z,x,\theta\,)$ is an operator whose
explicit form can be obtained from eq. \treinta. Using
this last equation one gets:
$$
\eqalign{
G_{j_1, j_2 }(\,z,x,\theta\,)\,=&\,{t\over 2}\,
(\,L_{-1}\,+\,\partial_z\,)\,-\,
(\,J_{-1}^+\,+\,z^{-1}\,\widetilde D^{\,+}_{j_1}\,)\,
(\,J_{0}^-\,-\,\widetilde D^{\,-}_{j_1}\,)\,-\,
2j_2\,(\,J_{-1}^0\,+\,z^{-1}\,\widetilde
D^{\,0}_{j_1}\,)\,+\cr
&+\,{1\over 2}\,
(\,j_{-1}^+\,+\,z^{-1}\,\widetilde d^{\,+}_{j_1}\,)\,
(\,j_{0}^-\,-\,\widetilde d^{\,-}_{j_1}\,)\,\,,
\cr}
\eqn\ctcsiete
$$
where the operators $\widetilde D^{\,a}_{j_1}$ and 
$\widetilde d^{\,\alpha}_{j_1}$ are obtained from those
of eq. \vseis\ by changing $x\rightarrow -x$ and
$\theta\rightarrow -\theta$, \ie\ they are given by:
$$
\eqalign{
\widetilde D^{\,0}_j\,=&\,-x\partial_x\,
-\,{1\over 2}\,\theta\,
\partial_{\theta}\,+\,j\cr
\widetilde D^{\,+}_j\,=&\,x^2\partial_x\,
-\,2jx\,+\,\theta x
\partial_{\theta}\cr
\widetilde D^{\,-}_j\,=&\,-\partial_x\cr
\widetilde d^{\,+}_j\,=&\,x\partial_{\theta}\,-\,\theta
x\partial_x\, +\,2j\theta\cr
\widetilde d^{\,-}_j\,=&\,-\partial_{\theta}\,
+\,\theta\partial_x\,\,.\cr}
\eqn\ctcocho
$$

On the other hand, it is proved in appendix C that the
two vectors $|\,\Lambda\,>$ and 
$|\,\widetilde \Lambda\,>$ defined as:
$$
\eqalign{
|\,\Lambda\,>\,=&\,\phi_{j_2}\,(z,x,\theta)\,
\,|\,j_1\,,\,t\,>\cr\cr
|\,\widetilde \Lambda\,>\,=&\,
e^{zL_{-1}\,+\,xJ_0^{-}\,+\,\theta j_{0}^-}\,
\phi_{j_1}\,(-z,-x,-\theta)\,
\,|\,j_2\,,\,t\,>\,\,,\cr}
\eqn\ctcnueve
$$
satisfy the same set of defining constraints and,
therefore, they can be identified. It is interesting at
this point to notice that the vector 
$|\,\widetilde\Lambda\,>$ can be generated in the
left-hand side of \ctcseis\ by inserting the exponential 
$e^{zL_{-1}\,+\,xJ_0^{-}\,+\,\theta j_{0}^-}$ and its
inverse. Using the 
$|\,\Lambda\,>\equiv |\,\widetilde \Lambda\,>$
identification, one arrives at:
$$
\eqalign{
e^{zL_{-1}\,+\,xJ_0^{-}\,+\,\theta j_{0}^-}\,
G_{j_1, j_2 }(\,z,x,\theta\,)\,
e^{-zL_{-1}\,-\,xJ_0^{-}\,-\,\theta j_{0}^-}\,
\phi_{j_2}\,(z,x,\theta)\,
\,|\,j_1\,,\,t\,>\,=\,0\,\,.\cr}
\eqn\ctcincuenta
$$
To proceed further with the calculation one has to
conjugate the operator $G_{j_1, j_2 }(\,z,x,\theta\,)$. The
conjugation of the currents with $e^{xJ_0^{-}\,+\,\theta
j_{0}^-}$ was given in eq. \vsiete. Moreover, the
behaviour of the derivatives \ctcocho\ under conjugation
is easy to obtain from their explicit expressions. After
a simple calculation one gets:
$$
\eqalign{
e^{\,xJ_0^{-}\,+\,\theta j_{0}^-}\,
\widetilde D^{\,0}_j\,
e^{\,-xJ_0^{-}\,-\,\theta j_{0}^-}\,=&\,
\widetilde D^{\,0}_j\,+\,xJ_{0}^-\,+\,
{1\over 2}\,\theta\,j_{0}^-\cr
e^{\,xJ_0^{-}\,+\,\theta j_{0}^-}\,
\widetilde D^{\,+}_j\,
e^{\,-xJ_0^{-}\,-\,\theta j_{0}^-}\,=&\,
\widetilde D^{\,+}_j\,-\,x^2J_{0}^-\,-\,
\,\theta x\,j_{0}^-\cr
e^{\,xJ_0^{-}\,+\,\theta j_{0}^-}\,
\widetilde D^{\,-}_j\,
e^{\,-xJ_0^{-}\,-\,\theta j_{0}^-}\,=&\,
\widetilde D^{\,-}_j\,+\,J_{0}^-\cr
e^{\,xJ_0^{-}\,+\,\theta j_{0}^-}\,
\widetilde d^{\,+}_j\,
e^{\,-xJ_0^{-}\,-\,\theta j_{0}^-}\,=&\,
\widetilde d^{\,+}_j\,+\,2\theta x\,J_0^{-}\,-\,
xj_0^{-}\cr
e^{\,xJ_0^{-}\,+\,\theta j_{0}^-}\,
\widetilde d^{\,-}_j\,
e^{\,-xJ_0^{-}\,-\,\theta j_{0}^-}\,=&\,
\widetilde d^{\,-}_j\,+\,j_{0}^-\,-\,
2\theta J_0^{-}\,\,.\cr}
\eqn\ctciuno
$$

In eq. \ctcincuenta\ one must also conjugate with the
operator $e^{zL_{-1}}$. This conjugation does not affect
the derivatives \ctcocho\ and the zero-mode currents
$J_0^{a}$ and $j_0^{\alpha}$. This  result is due
to the fact (see eq. \nueve) that 
$[L_{-1}, J_0^{a}]\,=\,[L_{-1},j_0^{\alpha}]\,=\,0$. 
Moreover, one can easily establish that:
$$
\eqalign{
e^{zL_{-1}}\,(\,L_{-1}\,+\,\partial_z\,)\,
e^{-zL_{-1}}\,=&\,\partial_z\cr
e^{zL_{-1}}\,\,J_{-1}^a\,\,e^{-zL_{-1}}\,=&\,
\sum_{p=1}^{+\infty}\,
J_{-p}^a\,\,z^{p-1}  \cr
e^{zL_{-1}}\,\,j_{-1}^{\alpha}\,\,e^{-zL_{-1}}\,=&\,
\sum_{p=1}^{+\infty}\,
j_{-p}^{\alpha}\,\,z^{p-1}\,\,.\cr}
\eqn\ctcidos
$$
On the other hand, let us define the  ``negative" part
of the currents as: 
$$
\eqalign{
&\widetilde J^{\,+}(z)\,\equiv\,\sum_{p=1}^{+\infty}\,
J_{-p}^+\,\,z^{p-1}
\,\,\,\,\,\,\,\,\,\,\,\,\,\,\,\,\,\,\,\,\,\,\,\,\,\,\,
\widetilde j^{\,+}(z)\,\equiv\,\sum_{p=1}^{+\infty}\,
j_{-p}^{+}\,\,z^{p-1}\cr
&\widetilde J^{\,-}(z)\,\equiv\,\sum_{p=0}^{+\infty}\,
J_{-p}^-\,\,z^{p-1}
\,\,\,\,\,\,\,\,\,\,\,\,\,\,\,\,\,\,\,\,\,\,\,\,\,\,\,
\widetilde j^{\,-}(z)\,\equiv\,\sum_{p=0}^{+\infty}\,
j_{-p}^{-}\,\,z^{p-1}\cr
&\widetilde J^{\,0}(z)\,\equiv\,\sum_{p=1}^{+\infty}\,
J_{-p}^0\,\,z^{p-1}\,\,.\cr}
\eqn\ctcitres
$$
Using these results, one can rewrite eq. \ctcincuenta\ as:
$$
\widehat G_{j_1,j_2}\,(\,z,x,\theta\,)
\phi_{j_2}\,(z,x,\theta)\,
\,|\,j_1\,,\,t\,>\,=\,0\,\,,
\eqn\ctcicuatro
$$
where $\widehat G_{j_1,j_2}\,(\,z,x,\theta\,)$ is the
operator:
$$
\eqalign{
\widehat G_{j_1,j_2}\,(\,z,x,\theta\,)\,=&\,
{t\over 2}\,\partial_z\,+\,
z^{-1}\,\Bigl[\,D_{j_2}^+\,D_{j_2}^-\,-\,
2j_1\,D_{j_2}^0\,-\,{1\over 2}\,
d_{j_2}^+\,d_{j_2}^-\,\Bigr]\,-
\widetilde J^{\,+}\,D_{j_2}^-\,-\,\cr
&-2\widetilde J^{\,0}\,D_{j_2}^0\,-\,
\widetilde J^{\,-}\,D_{j_2}^+\,+\,
{1\over 2}\,\,\widetilde j^{\,+}\,d_{j_2}^-\,-\,
{1\over 2}\,\,\widetilde j^{\,-}\,d_{j_2}^+\,\,.
\cr}
\eqn\ctcicinco
$$

Notice that the derivatives \vseis\ (and not those defined
in eq. \ctcocho) appear in 
$\widehat G_{j_1,j_2}\,(\,z,x,\theta\,)$. If we now
substitute the expansion of  $\phi_{j_2}\,(z,x,\theta)\,
\,|\,j_1\,,\,t\,>$ written in eqs. \ttres\ and \csiete\
and the expression \vseis\ of the derivatives, one can
easily obtain the recursion relations \ctvocho\ from eq.
\ctcicuatro. This is the result we wanted to demonstrate.

\chapter{Conclusions and outlook}

Let us now summarize our main results. We have been able
to set up a formalism in which the fusion of general 
${\rm osp}(1\vert 2)$ Verma module can be properly
defined. The isotopic dependence of the primary fields,
together with the differential realization \vseis\ of the
finite algebra, are the crucial ingredients  of our
approach. Using the $(x,\theta)$ dependence of the three
point function (eq. \scuatro), we have obtained a set of
singular vector decoupling conditions (eqs. \stcuatro\
and \stcinco) that encode the constraint induced in the
coupling of three Verma modules when one of them is
reducible. From these singular vector decoupling
conditions, we have found two set of fusion rules that the
product of primary operators corresponding to admissible
representations must obey. Moreover, we have obtained the
action of the currents of ${\cal A}_+$ on the fusion
states $|n,m>_{j_3}$ (\ie\ the descent equations \ctocho\
and \ctnueve). These  $|n,m>_{j_3}$ vectors satisfy a set
of recursion relations (eq. \ctvocho) that can be derived
either from the Sugawara energy-momentum tensor (as in
section 6) or, as was demonstrated in section 7, from
the Knizhnik-Zamolodchikov  equation.

At first sight our results are very similar to the ones
found for the $sl(2)$ current algebra [\AY, \Bauer]. It is
interesting to point out, however, the relevant r\^ole
played in our  ${\rm osp}(1\vert 2)$ case by the
Grassmann parity of the representations. Indeed, the
relative Grasssmann parity of the highest weight vectors
involved in the fusion appears explicitly in the
decoupling conditions \stcuatro\ and \stcinco, in the
descent equations \ctocho\ and \ctnueve, and in the
recursion relations \ctvocho. As a consequence, the
parity of the $[r_3,s_3]$ representation resulting from
the fusion of the two admissible representations 
$[r_1,s_1]$ and $[r_2,s_2]$ is fixed by eqs. \ndos\ and
\cttres\ and,  to obtain the singular vectors of
the algebra, the parameter $\delta_j$ must be carefully
adjusted as in eq. \cttnueve.

In order to construct a well-defined 
${\rm osp}(1\vert 2)$ CFT for general isospins and
levels, one should be able to define the conformal blocks
of the model. For the admissible ${\rm osp}(1\vert 2)$
representations, one expects that this could be done 
in the framework of a free field realization. This is
actually what occurs in the $sl(2)$ theory 
\REF\PRY{J. L. Petersen, J. Rasmussen and M. Yu
\journal\np&B481(96)577.}
[\peter, \PRY].
In the 
${\rm osp}(1\vert 2)$ case we have at our disposal all
the elements needed to define a free field realization of
the conformal blocks for the  correlators of
fields associated to admissible representations. In fact,
in the free field realization of 
${\rm osp}(1\vert 2)$, there exist two screening fields,
one of them is local in the free fields [\bershadsky],
while the other is non-local [\osp]. For general admissible
representations these two screening operators would be
needed to represent the blocks, while, as shown in ref.
[\osp], when
$k\in\ZZ_+$ and the isospins are integer or half-integer,
only the local screening is necessary. In order to deal
with the expectation values of non-local powers of the
fields that would appear in the correlators of the
general case, one can make use  of the fractional
calculus techniques [\Fractional], conveniently adapted, as
in eq.
\sseis, to include fermionic variables. 

One would expect to find in
this free field approach a more concrete operator
implementation of the quantum hamiltonian reduction. The
two fusion rules found in section 4 should also appear in
this formalism,   as it occurs for the $sl(2)$ case
[\PRY].  Moreover, it should be possible to characterize
completely the intermediate channels corresponding to both
set of fusion rules. In the approach of ref. [\PRY], the
second
$sl(2)$ fusion rule is obtained by the overscreening
mechanism, which is a consequence of the freedom that
exists, when the level is rational,  in the election of
the number of screening operators for a given correlator.
The Grassmann parity assignments of eqs. \ndos\ and
\cttres, together with the fact that the two ${\rm
osp}(1\vert 2)$ screening operators are fermionic, are a
hint which seems to indicate that, in our ${\rm
osp}(1\vert 2)$ case, the different screening
prescriptions could generate both types of fusion rules.
In order to get a definitive conclusion on this question
more work is needed. We expect to report on this matter
and on other related subjects in a near future.

\ack
We are grateful to J. M. Sanchez de Santos for
discussions and a critical reading of the manuscript. This
work was supported in part by DGICYT under grant
PB93-0344,  by CICYT under grant  AEN96-1673 and by the
European Union TMR grant ERBFMRXCT960012.

\Appendix A

In this appendix we shall study the characters of the 
${\rm osp}(1\vert 2)$  current algebra. For an 
irreducible Verma module whose highest weight vector has
isospin $j$, the character $\lambda_{j}(z,\tau)$ is
defined as:
$$
\lambda_{j}(z,\tau)\,=\,{\rm Tr}_j\,
[\,q^{L_0-{c\over 24}}\,w^{J_0^0}\,]\,,
\eqn\apauno
$$
where $q$ and $w$ are two variables related to the
modular parameter $\tau$ and to the coordinate $z$ by
means of the expressions:
$$
q\,=\,e^{2\pi i \tau}
\,\,\,\,\,\,\,\,\,\,\,\,\,\,\,\,\,\,\,\,
w\,=\,e^{2\pi i z}\,\,.
\eqn\apados
$$
The trace in \apauno\ must be taken over the module 
$V^{(j,t)}$. Its explicit expression can be obtained by
evaluating the action of the operator 
$q^{L_0-{c\over 24}}\,w^{J_0^0}$ on the states 
$|\,\{m_i^a\}\,;j\,>$, defined in section 2, 
that span $V^{(j,t)}$. Since 
$L_0$ and $J_0^0$ act diagonally on these states, the
trace \apauno\ can be easily computed. One gets the
following expression for $\lambda_{j}(z, \tau)$:
$$
\eqalign{
\lambda_{j}(z, \tau)\,=\,q^{h_j-{c\over 24}}\,
w^j\,
{\prod_{n=1}^{\infty}\,(1+q^n w^{{1\over 2}})\,
\prod_{n=0}^{\infty}\,(1+q^n w^{-{1\over 2}})\over
\prod_{n=1}^{\infty}\,(1-q^n )\,
\prod_{n=1}^{\infty}\,(1-q^n w)\,
\prod_{n=0}^{\infty}\,(1-q^n w^{-1})}\,\,.\cr\cr}
\eqn\apatres
$$
By using the identities:
$$
\eqalign{
{\prod_{n=1}^{\infty}\,(1+q^n w^{{1\over 2}})\over
\prod_{n=1}^{\infty}\,(1-q^n w)}\,=&\,
{1\over \prod_{n=1}^{\infty}\,(1-q^{n} w^{{1\over 2}})
\,(1-q^{2n-1} w)}\cr\cr
{\prod_{n=0}^{\infty}\,(1+q^n w^{-{1\over 2}})\over
\prod_{n=0}^{\infty}\,(1-q^n w^{-1})}\,=&\,
{1\over \prod_{n=1}^{\infty}\,(1-q^{n-1} w^{-{1\over 2}})
\,(1-q^{2n-1} w^{-1})}\,,\cr}
\eqn\apacuatro
$$
one can reexpress $\lambda_{j}(z, \tau)$ as:
$$
\lambda_{j}(z, \tau)\,=\,
{q^{ {2(j\,+{1\over 4})^2\over 
2k\,+\,3}}\,w^{j+\,{1\over 4}}
\over\Pi(z,\tau)}\,,
\eqn\apacinco
$$
where the function $\Pi(z,\tau)$ appearing in the
denominator of eq. \apacinco\ has the following infinite
product representation:
$$
\Pi(z, \tau)\,\equiv\,q^{{1\over 24}}\,w^{{1\over 4}}\,
\prod_{n=1}^{+\infty}\,\,
(1-q^n)\,(1-w^{{1\over 2}}q^n)\,(1-w^{-{1\over 2}}q^{n-1})
\,(1-wq^{2n-1})\,(1-w^{-1}q^{2n-1})\,.
\eqn\apaseis
$$
In the derivation of eq. \apacinco\ one must use that:
$$
h_j-{c-1\over 24}\,=\,{2(j\,+{1\over 4})^2\over 
2k\,+\,3}\,\,,
\eqn\apasiete
$$
which can be easily checked by a direct calculation from
eqs. \once\ and \quince. It is convenient in what follows
to rewrite  $\Pi(z, \tau)$ as an infinite sum. To achieve 
this objective it is enough to use the Watson quintuple
product identity, which reads:
$$
\eqalign{
\prod_{n=1}^{+\infty}\,\,&
(1-q^n)\,(1-wq^n)\,(1-w^{-1}q^{n-1})
\,(1-w^2q^{2n-1})\,(1-w^{-2}q^{2n-1})\,=\,\cr
&=\,\sum_{m=-\infty}^{+\infty}\,\,
(w^{3m}-w^{-3m-1})\,q^{{3m^2+m\over 2}}\,\,.\cr}
\eqn\apaocho
$$
Indeed, as a consequence of the identity \apaocho, one
can immediately verify that $\Pi(z, \tau)$ can be written
as:
$$
\Pi(z,\tau)\,=\,q^{{1\over 24}}\,w^{{1\over 4}}\,
\sum_{m\in\ZZ}\,\,
(w^{{3m\over 2}}-w^{-{3m+1\over 2}})\,
q^{{3m^2+m\over 2}}\,\,.
\eqn\apanueve
$$
The main consequence of eq. \apanueve\ is the fact that 
$\Pi(z,\tau)$ can be put as a difference of two classical 
theta functions. In general, the latter are defined as:
$$
\Theta_{r,s}\,(z,\tau)\,=\,
\sum_{m\in \ZZ}\,q^{s(m+{r\over 2s})^2}\,
w^{s(m+{r\over 2s})}\,.
\eqn\apadiez
$$
From this definition it follows by inspecting the
right-hand side of eq. \apanueve\ that:
$$
\Pi(z,\tau)\,=\,
\Theta_{1,3}\,( {z\over 2},{\tau\over 2})\,-\,
\Theta_{-1,3}\,({z\over 2},{\tau\over 2})\,\,.
\eqn\apaonce
$$
Let us now consider  the case of admissible
representations of the ${\rm osp}(1\vert 2)$  current
algebra. As  was discussed in the main text, these
representations appear when the level $k$ is such that 
$2k+3\,=\,{p\over p\,'}$, where $p$ and $p\,'$ are two
coprime integers and $p+p\,'\,\in 2\ZZ$. The isospins
$j_{r,s}$ corresponding to these representations are
labelled by two integers $r$ and $s$ which take values in
the grid $1\le r\le p-1$, $0\le s\le p\,'-1$ with $r+s\in
2\ZZ+1$. The actual values of $j_{r,s}$ are:
$$
4j_{r,s}\,+\,1\,=\,r\,-\,s\,{p\over p\,'}\,\,.
\eqn\apadoce
$$
When $j=j_{r,s}$, the Verma module $V^{(j,t)}$ is not
irreducible. The irreducible highest weight module for
these isospins is obtained by taking the quotient of 
$V^{(j,t)}$ by its maximum proper submodule. In fact, as
we have discussed in section 4, for $j=j_{r,s}$ the module
$V^{(j,t)}$ has two singular vectors with $J_0^0$
eigenvalues $j_{r,s}-{r\over 2}$ and  
$j_{r,s}-{r-p\over 2}$. These two vectors generate the
maximum proper submodule of  $V^{(j,t)}$, 
which can be represented by
means of the following embedding diagram:

$$
\def\sp{\nearrow\!\!\!\!\!\!\searrow}
\matrix{&&b(0)&\longrightarrow&a(1)&\longrightarrow&b(1)
&\longrightarrow&a(2)&\longrightarrow&\cdots\cr
a(0)&\nearrow\atop\searrow&&\sp&&\sp&&\sp&&\sp&\cdots\cr
&&b(-1)&\longrightarrow&a(-1)&\longrightarrow
&b(-2)&\longrightarrow&a(-2)
&\longrightarrow&\cdots\cr}
$$

\noindent
where $a(l)$ and $b(l)$ are given by:
$$
\eqalign{
a(l)\,&\equiv\,j_{r-2lp\,,\,s}\,=\,
{r-1\over 4}\,-\,{s\over 4}\,{p\over p\,'}\,-\,
l\,{p\over 2}\cr
b(l)\,&\equiv\,j_{-r-2lp\,,\,s}\,=\,
{r-1\over 4}\,-\,{s\over 4}\,{p\over p\,'}\,-\,
l\,{p\over 2}\,-\,{r\over 2}\,\,.\cr}
\eqn\apatrece
$$
Each node in the above diagram represents a Verma module
with $a(l)$ or $b(l)$ as the isospin of its highest
weight state. An arrow connecting two spaces 
$E\rightarrow F$ means that the module $F$ is contained
in the module $E$. The character of the irreducible
module with isospin $j=j_{r,s}$ is constructed as an
alternating sum of the form:
$$
\chi_{j_{r,s}}(z,\tau)\,=\,
\sum_{l=-\infty}^{l=+\infty}\,\,
\lambda_{a(l)}(z,\tau)\,-\,
\sum_{l=-\infty}^{l=+\infty}\,\,
\lambda_{b(l)}(z,\tau)\,\,.
\eqn\apacatorce
$$
Using eqs. \apacinco\ and \apatrece\ in the right-hand
side of eq. \apacatorce, it is straightforward to prove
that $\chi_{j_{r,s}}(z,\tau)$ can be written as a
quotient of differences of theta functions. Actually,
defining the constants $b_{\pm}$ and $a$ as:
$$
b_{\pm}\,=\,\pm p\,' r\,-\,p\,s
\,\,\,\,\,\,\,\,\,\,\,\,\,\,\,\,\,\,\,\,\,\,\,\,\,
a\,=\,p\,p\,'\,\,,
\eqn\apaquince
$$
the characters $\chi_{j_{r,s}}(z,\tau)$ can be put in the
form:
$$
\chi_{j_{r,s}}(z,\tau)\,=\,
{\Theta_{b_{+}, a}({z\over 2p\,'},{\tau\over 2}\,)\,-\,
\Theta_{b_{-}, a}( {z\over 2p\,'},{\tau\over 2}\,)
\over \Pi(z, \tau)}\,\,.
\eqn\apadseis
$$

We are interested in analyzing the $z=0$ behaviour of 
$\chi_{j_{r,s}}(z,\tau)$. It is easy to demonstrate that
the denominator function $\Pi(\tau,z)$ vanishes linearly
when  $z\rightarrow 0$. A simple calculation shows that:
$$
\Pi(z,\tau)\,=\,i\pi z q^{{1\over 24}}\,
\sum_{m\in \ZZ}\,(6m+1)\,q^{{3m^2+m\over 2}}\,+\,
o(z^2)\,\,.
\eqn\apadsiete
$$

We shall see below that, in general, the numerator of the
right-hand side of eq. \apadseis\ is non-vanishing. This
fact implies that $\chi_{j_{r,s}}(z,\tau)$ will, in
general, develop a simple pole in $z$ in the 
$z\rightarrow 0$ limit. The situation is very similar to
the one found in ref. [\Mukhi] for the admissible
representations of the $sl(2)$ current algebra. In this
latter case, a relation between the residues of the 
$sl(2)$ characters at the $z=0$ pole and the Virasoro
characters for the $c<1$ minimal models was found. In
our case, one would expect to find the characters of the
minimal supersymmetric models in the residue of 
$\chi_{j_{r,s}}(z,\tau)$ at $z=0$. This is actually what
happens, as we shall shortly prove. First of all, let us
rewrite the $z\rightarrow 0$ expansion \apadsiete\ in a
more convenient form. With this purpose in mind, let us
recall the infinite product representation of the
Dedekind $\eta$-function:
$$
\eta (\tau)\,=\,q^{{1\over 24}}\,
\prod_{n=1}^{\infty}\,(1\,-\,q^n)\,\,.
\eqn\apadocho
$$
Moreover, an identity due to Gordon 
\REF\Gordon{B. Gordon\journal\qjm&12(61)285.}
[\Gordon] allows to
express the sum appearing in the right-hand side of eq.
\apadsiete\ as an infinite product. In terms of 
$\eta(\tau)$ and the Jacobi theta function
$\theta_2(0,\tau)$, the Gordon identity can be written as:
$$
q^{{1\over 24}}\,
\sum_{m\in \ZZ}\,(6m+1)\,q^{{3m^2+m\over 2}}
\,=\,2\,
{\Big[\,\eta(\tau)\,\Big]^4\over
\theta_2(0,\tau)}\,\,.
\eqn\apadnueve
$$
The infinite product representation of $\theta_2(0,\tau)$ 
can be obtained from its relation with the
Dedekind function, namely:
$$
{\theta_2(0,\tau)\over \eta (\tau)}\,=\,
2\,q^{{1\over 12}}\,\prod_{n=1}^{\infty}\,
(\,1\,+\,q^{n}\,)^2\,\,.
\eqn\apaveinte
$$
Using eq. \apadnueve, one can write the $z\rightarrow 0$
expansion of $\Pi(z,\tau)$ as follows:
$$
\Pi(z,\tau)\,=\,2i\pi z\,\,
{\Big[\,\eta(\tau)\,\Big]^4\over
\theta_2(0,\tau)}
\,+\,o(z^2)\,\,.
\eqn\apavuno
$$
Let us now consider the difference of theta functions
appearing in the numerator of the right-hand side of eq.
\apadseis. Using the definition of the functions
$\Theta_{b_{\pm}, a}$ (see eq. \apadiez), one can write
this difference as:
$$
\eqalign{
&\Theta_{b_{+}, a}(\,{z\over 2p\,'}
\,,\,{\tau\over 2}\,)\,-\,
\Theta_{b_{-}, a}(\,{z\over 2p\,'}
\,,\,{\tau\over 2}\,)
\,=\,\cr\cr &=\,w^{{b_{+}\over
4p\,'}}\,q^{{(b_{+})^2\over 8a}}\,\,
\sum_{m\in\ZZ}\,q^{{am^2+mb_{+}\over 2}}\,\,
\Big(\,w^{{am\over 2p\,'}}\,-\,q^{{s\over 2}(r+2p\,'tm)}\,
w^{-{am\over 2p\,'}-{r\over 2}}\,\Big)\,\,.
\cr}
\eqn\apavdos
$$
Taking $z=0$ in eq. \apavdos, one gets:
$$
\Theta_{b_{+}, a}(\,0,{\tau\over 2}\,)\,-\,
\Theta_{b_{-}, a}(\,0,{\tau\over 2}\,)\,
=\,\sum_{m\in\ZZ}\,\Big[\,
q^{{(\lambda+2pp\,'m)^2\over 8pp\,'}}\,\,-\,\,
q^{{(\bar\lambda+2pp\,'m)^2\over 8pp\,'}}\,\,\Big]\,\,,
\eqn\apavtres
$$
where $\lambda$ and $\bar\lambda$ are given by:
$$
\lambda\,=\,ps\,-\,p\,'r
\,\,\,\,\,\,\,\,\,\,\,\,\,\,\,\,\,\,\,\,\,\,\,\,\,
\bar\lambda\,=\,ps\,+\,p\,'r\,\,.
\eqn\apavcuatro
$$
When $s\not= 0$, the right-hand side of eq. \apavtres\ is
non-vanishing and, therefore, it makes sense to consider
the residue of $\chi_{j_{r,s}}(z,\tau)$ at the point
$z=0$. Let us define in this case the following quantity:
$$
\hat\chi_{r,s}(\tau)\,\equiv\,
\Big[\,{2\eta(\tau)\over\theta_2(0,\tau)}\,
\Big]^{1\over 2}\,\,
\Big[\,\eta(\tau)\,\Big]^2\,\,
{\rm lim}_{z\rightarrow 0}\,\,
\Big\{\,i\pi z\,
\chi_{j_{r,s}}(z,\tau)
\,\Big\}\,\,.
\eqn\apavcinco
$$
As a consequence of our previous results (eqs. \apavuno\
and \apavtres), $\hat\chi_{r,s}(\tau)$ is equal to:
$$
\hat\chi_{r,s}(\tau)\,=\,
\Big[\,{\theta_2(0,\tau)
\over 2\eta(\tau)}\,\Big]^{1\over 2}\,\,\,\,
{\Theta_{b_{+}, a}(\,0,{\tau\over 2}\,)\,-\,
\Theta_{b_{-}, a}(\,0,{\tau\over 2}\,)\over
\eta(\tau)}\,\,.
\eqn\apavseis
$$
It is interesting to point out that for 
$1\le r\le p-1\,\,\,$, $1\le s\le p\,'-1\,\,$  and
$r+s\in 2\ZZ+1$, the functions of $\tau$ appearing in
the right-hand side of eq. \apavseis\ are precisely the
characters of the minimal supersymmetric models, with
central charge 
$c\,=\,{3\over 2}\,(1\,-\,{2(p-p\,')^2\over pp\,'})$, in the
Ramond sector. This is precisely the result we were
looking for.

\Appendix B

In this appendix we shall illustrate the algorithm 
to compute singular vectors 
described in section 6 by performing the explicit
calculation of the simplest non-trivial cases. Let us,
first of all, consider the case $r=1$ and $s=2$. The
corresponding isospin is (see eq. \cocho) $j_{1,2}=-t/2$.
According to our general prescription we must take 
$i_-\,=\,-1/2$ and $i_+\,=\,1/2$ in the descent
equations. For these values of $i_{\pm}$, the numbers $n$
and $m$ appearing in the Sugawara recursion relations are
restricted to the ranges 
$\,\,0\,\le\,n\,\le1\,\,$ and $-1/2\,\le m\,\le
1/2\,$ (see eq. \ctcdos) and the vector
$|\,\lambda_{1,2}\,>$ is given by:

$$
\eqalign{
|\,\lambda_{1,2}\,>\,=&\,{1\over 2}
\,j_{-1}^-|\,0,0\,>_{j_{1,2}}\,+\,
J_{-1}^0\,|\,0,{1\over 2}\,>_{j_{1,2}}\,+\cr
&+\,J_{0}^-\,|\,1,-{1\over 2}\,>_{j_{1,2}}
\,+\,{1\over 2}\,j_{0}^-|\,1,0\,>_{j_{1,2}}\,\,.\cr}
\eqn\apbuno
$$

In order to obtain the vectors $|\,n,m\,>_{j_{1,2}}$ 
appearing in the right-hand side of eq. \apbuno, one
must solve the recursion relations \ctvocho. When 
$j_3\,=\,j_{1,2}\,=\,-t/2$, eq. \ctvocho\  for $m=0, \pm
1/2$ gives rise to the following equations:
$$
\eqalign{
{1\over 2}\,[\,(n+1)\,t\,-\,1\,]\,
|\,n,-{1\over 2}\,>_{j_{1,2}}\,=&\,
\sum_{p=1}^{n}\,J_{-p}^+\,
|\,n-p,{1\over 2}\,>_{j_{1,2}}\,+\,
\sum_{p=1}^{n}\,J_{-p}^0\,
|\,n-p,-{1\over 2}\,>_{j_{1,2}}\,-\,\cr
&-\,{1\over 2}\,\sum_{p=1}^{n}\,j_{-p}^+\,
|\,n-p,0\,>_{j_{1,2}}\cr
{t\over 2}\,n\,|\,n,0\,>_{j_{1,2}}\,=&\,
{1\over 2}\,\sum_{p=1}^{n}\,j_{-p}^+\,
|\,n-p,\,{1\over 2}\,>_{j_{1,2}}\,+\,
{1\over 2}\,\sum_{p=0}^{n}\,j_{-p}^-\,
|\,n-p,\,-{1\over 2}\,>_{j_{1,2}}\cr
{t\over 2}\,(n-1)\,|\,n,\,{1\over 2}\,>_{j_{1,2}}\,=&\,
-\sum_{p=1}^{n}\,J_{-p}^0\,
|\,n-p,{1\over 2}\,>_{j_{1,2}}\,+\,
\sum_{p=0}^{n}\,J_{-p}^-\,
|\,n-p,\,-{1\over 2}\,>_{j_{1,2}}\,+\cr
&+\,{1\over 2}\,\sum_{p=0}^{n}\,j_{-p}^-\,
|\,n-p,0\,>_{j_{1,2}}\,\,.\cr}
\eqn\apbdos
$$
It is a simple exercise to solve \apbdos\ recursively.
One gets:
$$
\eqalign{
|\,0\,,\,{1\over 2}\,>_{j_{1,2}}\,=&\,-{1\over t}\,
j_0^-|\,0\,,0\,>_{j_{1,2}}\cr
|\,1\,,\,-{1\over 2}\,>_{j_{1,2}}\,=&\,
{1\over (1-2t)}\,\Bigl[\,{2\over t}J_{-1}^+j_0^-\,+\,
j_{-1}^+\,\Bigr]\,
|\,0\,,0\,>_{j_{1,2}}\cr
|\,1\,,\,0\,>_{j_{1,2}}\,=&\,-{1\over t^2}\,
\Bigl[\,j_{-1}^+j_{0}^-\,+\,{1\over 2t-1}\,
(\,2j_0^-J_{-1}^+j_{0}^-\,+\,tj_0^-j_{-1}^+\,)
\,\Bigr]\,|\,0\,,0\,>_{j_{1,2}}\,\,.
\cr\cr}
\eqn\apbtres
$$
Taking $|\,0,0\,>_{j_{1,2}}\,=\,|\,j_{1,2}, t\,>$ and
substituting eq. \apbtres\ in eq. \apbuno, we arrive
at:
$$
\eqalign{
|\,\lambda_{1,2}\,>\,=\,-{1\over t^2}\,
\Bigl[\,J_0^-J_{-1}^+j_0^-\,+\,(1-t)J_{-1}^0j_0^-
\,+\,{1+t\over 2}\,J_{0}^{-}j_{-1}^+\,+\,{1\over 2}\,
(1-t^2)j_{-1}^-\,\Bigr]\,|\,j_{1,2}\,,t\,>\,\,.\cr\cr}
\eqn\apbcuatro
$$
It can be checked directly that the vector
\apbcuatro\ satisfies all the conditions required to an
${\rm osp}(1\vert 2)$  singular vector. Moreover, up
to a constant, the vector \apbcuatro\ coincides with the
one obtained from the MFF expression (eq. \cincuenta).

In the same way, one could work out the $r=2$, $s=1$
case. We shall limit ourselves here to write the final
result for the corresponding vector  
$|\,\lambda_{2,1}\,>$:

$$
\eqalign{
|\,\lambda_{2,1}\,>\,=&\,{4\over t^2-1}\,\Bigl[\,
J_{-1}^+(J_0^-)^2\,-\,j_{-1}^+j_0^-J_0^-\,-\,
(t+1)J_{-1}^0J_0^{-}\,+\cr
&+\,{1\over 2}(t+1)j_{-1}^-j_0^{-}\,-\,
{1\over 4}\,(t^2-1)\,J_{-1}^-\,
\Bigr]\,|\,j_{2,1}\,,t\,>\,\,.\cr}
\eqn\apbcinco
$$

\Appendix C

In our derivation of the descent equations from the 
Knizhnik-Zamolodchikov equations in section 7 we have 
used the fact that the vectors $|\,\Lambda\,>$ and 
$|\,\widetilde \Lambda\,>$, defined in eq. \ctcnueve, can
be identified. In this appendix we shall verify that both
states satisfy the same set of covariance constraints
and,  therefore,  they should be consider as  identical.

The non-trivial constraints satisfied by the state
$|\,\Lambda\,>$ can be obtained from those verified by the
highest weight $|\,j_1\,,\,t\,>$. The latter are
determined from the left ideal of ${\cal A}$ annihilating 
$|\,j_1\,,\,t\,>$, namely:
$$
\eqalign{
(\,J_0^0\,-\,j_1\,)\,|\,j_1\,,\,t\,>\,=&\,
(\,L_0\,-\,j_1\,)\,|\,j_1\,,\,t\,>\,=\,0\cr
J_n^{a}\,|\,j_1\,,\,t\,>\,=&\,
j_n^{\alpha}\,|\,j_1\,,\,t\,>\,=\,0,
\,\,\,\,\,\,\,\,\,\,\,\,\,\,\,\,\,\,
\forall\,\,(\,J_n^a\,,\,j_n^{\alpha}\,)\,\in\,
{\cal A}_{+}\,\,.\cr}
\eqn\apcuno
$$
Multiplying by $\phi_{j_2}(z,x,\theta)$ the equations in
\apcuno\ and commuting it with the
operators that multiply the state $|\,j_1\,,\,t\,>$ in the
constraints \apcuno, we get the following conditions
satisfied by $|\,\Lambda\,>$:
$$
\eqalign{
(\,J_0^0\,-\,D_{j_2}^0\,-\,j_1\,)\,|\,\Lambda\,>\,=&\,
(\,L_0\,-\,z\partial_z\,-\,h_1\,-\,h_2\,)\,|\,\Lambda\,>\,=\,0\cr
(\,J_n^{a}\,-\,z^n\,D_{j_2}^a\,)\,|\,\Lambda\,>\,=&\,
(\,j_n^{\alpha}\,-\,z^n\,d_{j_2}^{\alpha}\,)\,|\,\Lambda\,>\,=0
\,\,\,\,\,\,\,\,\,\,\,\,\,\,\,\,\,\,
\forall\,\,(\,J_n^a\,,\,j_n^{\alpha}\,)\,\in\,
{\cal A}_{+}\,\,.
\cr\cr}
\eqn\apcdos
$$

We are going to prove that $|\,\widetilde \Lambda\,>$ also
satisfies eq. \apcdos. Actually, the constraints verified
by $|\,\widetilde \Lambda\,>$ can be derived following
steps similar to those employed to obtain \apcdos. One
starts, in this case, from the constraints satisfied by 
$|\,j_2\,,\,t\,>$ and multiplies them by 
$\phi_{j_1}(-z,-x,-\theta)$. After commuting this field
with the operators that annihilate $|\,j_2\,,\,t\,>$  and
performing a conjugation with 
$e^{zL_{-1}+xJ_0^-+\theta j_0^-}$, one gets:
$$
\eqalign{
e^{zL_{-1}+xJ_0^-+\theta j_0^-}\,(\,J_0^0\,-\,\widetilde
D_{j_1}^{\,0}\,-\,j_2\,)\,
e^{-zL_{-1}-xJ_0^--\theta j_0^-}\,\,
|\,\widetilde \Lambda\,>\,=&\,0\cr
e^{zL_{-1}+xJ_0^-+\theta j_0^-}\,(\,L_0\,
-\,z\partial_z\,-\,h_1\,-\,h_2\,)\,
e^{-zL_{-1}-xJ_0^--\theta j_0^-}\,\,
|\,\widetilde \Lambda\,>\,=&\,0\cr
e^{zL_{-1}+xJ_0^-+\theta j_0^-}\,(\,J_n^a\,-\,(-1)^n\,
z^n\,\widetilde D_{j_1}^{\,a}\,)\,
e^{-zL_{-1}-xJ_0^--\theta j_0^-}\,\,
|\,\widetilde \Lambda\,>\,=&\,0\cr
e^{zL_{-1}+xJ_0^-+\theta j_0^-}\,(\,j_n^{\alpha}
\,-\,(-1)^n\, z^n\,\widetilde d_{j_1}^{\,\alpha}\,)\,
e^{-zL_{-1}-xJ_0^--\theta j_0^-}\,\,
|\,\widetilde \Lambda\,>\,=&\,0\cr
&\forall\,\,(\,J_n^a\,,\,j_n^{\alpha}\,)\,\in\,
{\cal A}_{+}\,\,.\cr\cr}
\eqn\apctres
$$

In eq. \apctres, the operators $\widetilde D_{j}^{\,a}$ and
$\widetilde d_{j}^{\,\alpha}$ are the ones defined in
section 7 (see eq. \ctcocho). We claim that eq. \apctres\
implies eq. \apcdos\ with $|\,\Lambda\,>$ substituted by 
$|\,\widetilde \Lambda\,>$. Let us consider, first of all,
the $J_0^0$ constraint. Using the conjugation properties
of $J_0^0$ (section 2, eq. \vsiete) and 
$\widetilde D_{j_1}^{\,0}$ (section 7, eq. \ctciuno), one
can convert the first eq. in \apctres\ into:
$$
(\,J_0^0\,-\,\widetilde D_{j_1}^0\,-\,j_2\,)\,
|\,\widetilde \Lambda\,>\,=\,0\,\,.
\eqn\apccuatro
$$
As $\widetilde D_{j_1}^0\,+\,j_2\,=\,D_{j_2}^0\,+\,j_1$,
eq. \apccuatro\ can be written as:
$$
(\,J_0^0\,-\,D_{j_2}^0\,-\,j_1\,)\,
|\,\widetilde \Lambda\,>\,=\,0\,\,,
\eqn\apccinco
$$
which is the first equation \apcdos\ with 
$|\,\widetilde \Lambda\,>$ instead of 
$|\,\Lambda\,>$, as claimed. Similarly we can demonstrate
that the  $L_0$ constraints satisfied by 
$|\, \Lambda\,>$ and $|\,\widetilde \Lambda\,>$ are the
same. Indeed, as $L_0\,-\,z\partial_z$ is invariant under
conjugation, one has:
$$
(\,L_0\,-\,z\partial_z\,-\,h_1\,-\,h_2\,)\,
|\,\widetilde \Lambda\,>\,=\,0
\,\,.
\eqn\apcseis
$$

Let us consider from now on the constraints induced by the
elements of ${\cal A}_+$. First of all, we notice that,
when $n\ge 0$, the result of conjugating with $L_{-1}$ the
currents $J_n^{a}$ and $j_n^{\alpha}$ is:
$$
\eqalign{
e^{zL_{-1}}\,J_n^{a}\,e^{-zL_{-1}}\,=&\,
\sum_{p=0}^{n}\,(-1)^p\,z^p\,
{n\choose p}J_{n-p}^{a}\cr
e^{zL_{-1}}\,j_n^{\alpha}\,e^{-zL_{-1}}\,=&\,
\sum_{p=0}^{n}\,(-1)^p\,z^p\,
{n\choose p}j_{n-p}^{\alpha}\,\,.\cr}
\eqn\apcsiete
$$
Using eqs. \apcsiete, \vsiete\ and \ctciuno, the $J_n^{-}$
constraint in \apctres\ can be written as:
$$
\sum_{p=0}^{n-1}\,(-1)^p\,z^p\,
{n\choose p}J_{n-p}^{-}\,|\,\widetilde \Lambda\,>\,+\,
(-1)^n\,z^n\,D_{j_2}^{-}\,|\,\widetilde \Lambda\,>\,
=\,0\,\,.
\eqn\apcocho
$$
Putting $n=1$ in eq. \apcocho, we get 
$(J_1^-\,-\,zD_{j_2}^-\,)\,|\,\widetilde
\Lambda\,>\,=\,0$, which implies that
$|\,\widetilde \Lambda\,>$ and $|\, \Lambda\,>$
satisfy the same $J_n^{-}$ 
constraint for $n=1$. It is not difficult to generalize
this result for an arbitrary value of $n$. Let us apply
the induction method and suppose that 
$|\,\widetilde \Lambda\,>$ satisfies:
$$
J_{n-p}^{-}\,|\,\widetilde \Lambda\,>\,=\,
z^{n-p}\,D_{j_2}^{-}\,|\,\widetilde \Lambda\,>\,,
\,\,\,\,\,\,\,\,{\rm for}\,\,1<p<n\,.
\eqn\apcnueve
$$
Making use of eq. \apcnueve\ in eq. \apcocho\ one gets:
$$
\Bigl[\,\sum_{p=1}^{n-1}\,(-1)^p\,{n\choose p}\,\Bigr]
\,z^n\,D_{j_2}^{-}\,|\,\widetilde \Lambda\,>\,+\,
J_{n}^{-}\,|\,\widetilde \Lambda\,>\,+\,
(-1)^n\,z^n\,D_{j_2}^{-}\,
|\,\widetilde \Lambda\,>\,=\,0\,.
\eqn\apcdiez
$$
As the sum in $p$ in eq. \apcdiez\ is $-1-(-1)^n$, it
follows that $|\,\widetilde \Lambda\,>$ verifies:
$$
(\,J_n^-\,-\,z^n\,D_{j_2}^-\,)\,
|\,\widetilde \Lambda\,>\,=\,0\,\,,
\,\,\,\,\,\,\,\,\,\,\,\,\,\,\,\,\,
(\,n\,\ge 1\,)\,\,,
\eqn\apconce
$$
which is the result we wanted to demonstrate. 

One can proceed similarly with the other currents. To
illustrate the procedure let us consider in detail the
case of $j_n^-$. Using eqs. \apcsiete, \vsiete\ and
\ctciuno\ in eq.\apctres, it is straightforward to prove
that:
$$
\sum_{p=0}^{n-1}\,(-1)^p\,z^p\,
{n\choose p}\,(\,j_{n-p}^{-}\,-\,2\theta J_{n-p}^{-}\,)\,
|\,\widetilde \Lambda\,>\,-\,(-1)^n\,z^n\,
\widetilde d_{j_1}^{\,-}\,|\,\widetilde \Lambda\,>\,=\,0
\,\,.
\eqn\apcdoce
$$
For $n=1$ eq. \apcdoce\ reduces to:
$$
(\,j_1^-\,-\,2\theta\,J_1^-\,+\,z\widetilde 
d_{j_1}^{\,-}\,)\,|\,\widetilde \Lambda\,>\,=\,0\,\,,
\eqn\apctrece
$$
which, taking into account that 
$J_1^-\,|\,\widetilde \Lambda\,>\,=\,
z\,D_{j_2}^{\,-}\,|\,\widetilde \Lambda\,>$
(see eq. \apconce), is equivalent to:
$$
(\,j_1^-\,-\,2\theta\,z\,D_{j_2}^{\,-}   
\,+\,z\widetilde  d_{j_1}^{\,-}\,)\,|\,\widetilde
\Lambda\,>\,=\,0\,\,.
\eqn\apccatorce
$$
An easy calculation, using the explicit expressions of the
differential operators, shows that 
$2\theta\,D_{j_2}^{\,-}   
\,-\,\widetilde  d_{j_1}^{\,-}\,=\,d_{j_2}^{\,-}$.
Substituting this result in eq. \apccatorce,  one gets:
$$
(\,j_1^-\,-\,z  d_{j_2}^{\,-}\,)\,|\,\widetilde
\Lambda\,>\,=\,0\,\,,
\eqn\apcquince
$$
which means that $|\,\Lambda\,>$ and
$|\,\widetilde\Lambda\,>$ satisfy the same $j_1^-$
constraint. To extend this result for arbitrary $n$, we
proceed, as before, by induction. Assuming that the
following equation holds:
$$
j_{n-p}^{-}\,|\,\widetilde \Lambda\,>\,=\,
z^{n-p}\,d_{j_2}^{-}\,|\,\widetilde \Lambda\,>\,
\,\,\,\,\,\,\,\,{\rm for}\,\,1<p<n\,\,,
\eqn\apcdseis
$$
together with eq. \apconce, the general $j_n^-$
constraint (eq. \apcdoce)  can be reduced to the form:
$$
j_{n}^{-}\,|\,\widetilde \Lambda\,>\,-\,
(-1)^n\,z^n\,(\,\widetilde d_{j_2}^{\,-}\,-\,
2\theta\,D_{j_2}^{-}\,+\,\widetilde d_{j_1}^{\,-}\,)
\,|\,\widetilde \Lambda\,>\,-\,z^n\,
 d_{j_1}^{\,-}\,|\,\widetilde \Lambda\,>\,=\,0\,\,.
\eqn\apcdsiete
$$
As $\widetilde d_{j_2}^{\,-}\,-\,
2\theta\,D_{j_2}^{-}\,+\,\widetilde d_{j_1}^{\,-}\,=\,0$,
it follows from eq. \apcdsiete\ that:
$$
(\,j_n^-\,-\,z^n d_{j_2}^{\,-}\,)\,|\,\widetilde
\Lambda\,>\,=\,0\,\,,
\,\,\,\,\,\,\,\,\,\,\,\,\,\,\,\,\,
(\,n\,\ge 1\,)\,\,.
\eqn\apcdocho
$$

Using similar arguments one can demonstrate that 
$|\,\widetilde\Lambda\,>$ satisfies:
$$
\eqalign{
(\,J_n^0\,-\,z^n D_{j_2}^{\,0}\,)\,|\,\widetilde
\Lambda\,>\,=&\,0
\,\,\,\,\,\,\,\,\,\,\,\,\,\,\,\,\,
(\,n\,\ge 1\,)\cr
(\,J_n^+\,-\,z^n D_{j_2}^{\,+}\,)\,|\,\widetilde
\Lambda\,>\,=&\,0
\,\,\,\,\,\,\,\,\,\,\,\,\,\,\,\,\,
(\,n\,\ge 0\,)\cr
(\,j_n^+\,-\,z^n d_{j_2}^{\,+}\,)\,|\,\widetilde
\Lambda\,>\,=&\,0
\,\,\,\,\,\,\,\,\,\,\,\,\,\,\,\,\,
(\,n\,\ge 0\,)\,\,.\cr}
\eqn\apcdnueve
$$
This completes the proof of the equivalence of 
$|\,\Lambda\,>$ and $|\,\widetilde\Lambda\,>$.

\refout

\end